\documentclass[pre,twocolumn,amsmath,amssymb,showpacs]{revtex4}
\usepackage{graphicx}
\usepackage{dcolumn}
\usepackage{bm}
\begin{document}
\newcommand{\vspfigA}{\vspace{0.45cm}
   \rule{85mm}{0.3mm}
   \vspace{-0.15cm}}  
\newcommand{\vspfigB}{\vspace{-0.05cm}
   \rule{85mm}{0.3mm}
   \vspace{0.35cm}} 
\newcommand{\vspfigAa}{\vspace{0.45cm}}  
\newcommand{\vspfigBa}{\vspace{0.35cm}}
\newcommand{\widthfigA}{0.37\textwidth} 
\newcommand{\widthfigB}{0.85\textwidth} 
\newcommand{\widthtableA}{2cm} 
\newcommand{\widthtableB}{1.8cm} 
\newcommand{\widthtableC}{0.1cm} 
\newcommand{\widthtableD}{\hspace{0.2cm}} 
\newcommand{\bfGamma}{\mbox{\boldmath $\bf\Gamma$}}
\baselineskip 0.405cm
%
%
\title{Time-dependent mode structure for Lyapunov vectors 
as a collective movement in quasi-one-dimensional 
systems}   
\author{Tooru Taniguchi and Gary P. Morriss}
\affiliation{School of Physics, University of New South Wales, 
Sydney, New South Wales 2052, Australia}
\date{\today}
\begin{abstract}
   Time dependent mode structure for the Lyapunov vectors 
associated with the stepwise structure of the Lyapunov spectra 
and its relation to the momentum auto-correlation function are 
discussed in quasi-one-dimensional many-hard-disk systems. 
   We demonstrate mode structures (Lyapunov modes) 
for all components of the 
Lyapunov vectors, which include the longitudinal and transverse 
components of their spatial and momentum parts, and their phase 
relations are specified. 
   These mode structures are suggested from the form 
of the Lyapunov vectors corresponding to the zero-Lyapunov exponents. 
   Spatial node structures of these modes are explained by 
the reflection properties of the hard-walls used in the models. 
   Our main interest is the time-oscillating behavior of Lyapunov 
modes. 
   It is shown that the largest time-oscillating period 
of the Lyapunov modes is twice as 
long as the time-oscillating period of 
the longitudinal momentum auto-correlation function. 
   This relation is satisfied irrespective of the particle number 
and boundary conditions.  
   A simple explanation for this relation is given based on 
the form of the Lyapunov vector. 
\end{abstract}
%
\vspace{1cm}
\maketitle
%
%
\section{Introduction}

   Statistical mechanics based on  
dynamical instability has drawn considerable attention 
in recent years. 
   The dynamical instability is described as a rapid separation 
of two nearby trajectories, the so called Lyapunov vector, 
and causes the loss of memory or unpredictability of the dynamical 
system. 
   The exponential rate of expansion or contraction of the 
magnitude of the Lyapunov vector is called the Lyapunov exponent, 
and its positivity, meaning that the system is chaotic, is a well 
known indicator of the dynamical instability. 
   Many efforts have been devoted to connect the dynamical 
instability with statistical properties, like transport 
coefficients \cite{Gas98,Dor99}. 
   Some works concentrated on specific effects of the dynamical 
instability in many-particle systems. 
   Information on the dynamical instability 
in many-particle systems is given by the set of Lyapunov 
exponents (the Lyapunov spectrum) and their associated Lyapunov 
vectors, and their structures have been 
of much interest to the study of chaotic properties of many-particle 
systems.  
   The conjugate pairing rule of Lyapunov spectra 
for thermostated systems \cite{Dre88,Eva90,Det96a,Tan02a}, 
localized behaviors for Lyapunov vectors 
\cite{Kan86,Liv89,Fal91,Mil02,Tan03b}, and 
the thermodynamic limit of Lyapunov spectra 
\cite{Pal86,New86,Sin96,Tan02b} and so on  
have been discussed from this point of view. 

   The stepwise structure of Lyapunov spectra is a typical 
chaotic property of many-particle systems, which 
was found recently \cite{Del96}.
   This stepwise structure appears in the Lyapunov 
exponents with smallest absolute value, and the dynamical structure 
of these Lyapunov exponents 
should reflect a slow and global behavior 
of the macroscopic system. 
   Therefore, clarification of the stepwise structure 
of Lyapunov spectra (Lyapunov steps) is expected to make 
a bridge between 
the macroscopic statistical theory and the microscopic 
chaotic dynamics. 
   Remarkably, the Lyapunov steps accompany 
wavelike structures in the associated Lyapunov vectors, 
namely the Lyapunov modes, which offer a useful tool 
to understand the origin of the stepwise structure 
of the Lyapunov spectrum \cite{Pos00,Tan03a,For04,Eck04}. 
   Originally, these structures were observed 
in many-hard-disk systems, but very recently 
numerical evidence for the Lyapunov modes was reported 
for many-particle system with soft-core particle interactions 
\cite{Yan04}.  
   Some theoretical arguments have been proposed to 
explain this phenomenon, for example, using random matrix 
theory \cite{Eck00,Tan02c}, kinetic theory 
\cite{Mcn01b,Mar04,Wij03}, and 
periodic orbit theory \cite{Tan02b}, etc.

   The first step to understand the Lyapunov step and mode 
is in the zero Lyapunov exponents and 
the associated Lyapunov vectors.    
   For instance, the two-dimensional system 
consisting of $N$ particles with periodic boundary conditions 
has six zero-Lyapunov exponents, and 
the associated Lyapunov vectors are linear combinations 
of the six basis vectors
%
   $N^{-1/2} 
       (\mathbf{1},\mathbf{0}, \mathbf{0},\mathbf{0})$,
   $N^{-1/2} 
       (\mathbf{0},\mathbf{1}, \mathbf{0},\mathbf{0})$,
   $N^{-1/2} 
       (\mathbf{0},\mathbf{0}, \mathbf{1},\mathbf{0})$,
   $N^{-1/2} 
      (\mathbf{0},\mathbf{0}, \mathbf{0},\mathbf{1})$,
   $|\mathbf{p}|^{-1} 
      (\mathbf{p}_{x},\mathbf{p}_{y}, \mathbf{0},\mathbf{0})$, and 
   $|\mathbf{p}|^{-1}
      (\mathbf{0},\mathbf{0}, \mathbf{p}_{x},\mathbf{p}_{y})$ 
%
where $\mathbf{0}$ is a $N$-dimensional null vector, 
$\mathbf{1}$ is an $N$-dimensional 
vector with all $x$-component equal to $1$, 
and $\mathbf{p}\equiv (\mathbf{p}_{x},\mathbf{p}_{y})$ is 
the momentum vector with its $x$-component $\mathbf{p}_{x}$ 
and $y$-component $\mathbf{p}_{y}$. 
   Here, the first (second) vector is associated 
with the translational invariance in the $x$-direction 
($y$-direction), the third (fourth) vector 
with the conservation of the $x$-component 
($y$-component) of the total momentum, the fifth vector 
with the time-translational invariance 
(deterministic nature of the orbit), and 
the last vector with the energy conservation \cite{Gas98}. 
   This means that the six sets of the Lyapunov vector 
components 
%
   $\{\delta x_{j}^{(n)} \}_{j}$,
   $\{\delta y_{j}^{(n)} \}_{j}$, 
   $\{\delta p_{xj}^{(n)} \}_{j}$, 
   $\{\delta p_{yj}^{(n)} \}_{j}$, 
   $\{\delta x_{j}^{(n)}/p_{xj},  
      \delta y_{j}^{(n)}/p_{yj} \}_{j}$, and 
   $\{\delta p_{xj}^{(n)}/p_{xj}, 
      \delta p_{yj}^{(n)}/p_{yj} \}_{j}$
%
can have equal components independent of the particle 
number index $j$ in zero-Lyapunov exponents. 
   Here, we use the notation $\delta\bfGamma^{(n)}
= (\delta\mathbf{q}^{(n)}, \delta\mathbf{p}^{(n)})$ 
for the Lyapunov vector corresponding to Lyapunov exponents 
$\lambda^{(n)}$, and 
$\delta\mathbf{q}^{(n)} 
\equiv (\delta x_{1}^{(n)}, \delta x_{2}^{(n)},$ 
$\cdots, \delta x_{N}^{(n)}, $ $
\delta y_{1}^{(n)}, \delta y_{2}^{(n)},
\cdots,\delta y_{N}^{(n)})$ and 
$\delta\mathbf{p}^{(n)}
\equiv (\delta p_{x1}^{(n)}, \delta p_{x2}^{(n)},
\cdots,\delta p_{xN}^{(n)}, $ $
\delta p_{y1}^{(n)}, \delta p_{y2}^{(n)},$ 
$\cdots,\delta p_{yN}^{(n)})$ 
are the spatial part and the momentum 
part of the Lyapunov vector, respectively. 

   As the next step we regard the degeneracy of the zero-Lyapunov 
exponents and the structure of the corresponding Lyapunov vectors 
as the zero-th Lyapunov step and mode. 
   This scenario was proposed first in Ref. \cite{Tan03a}, 
and was also discussed very recently in Ref. \cite{Eck04}. 
   First of all, the Lyapunov steps for the 
two-dimensional system with periodic boundary conditions 
consist of two-point steps and four-point steps \cite{Pos00,Tan03a}, 
namely the number of Lyapunov exponents for one set of 
Lyapunov steps is six, which is equal to the number 
of the zero-Lyapunov exponents. 
   It is also known that the stepwise structure 
of Lyapunov spectra is changed 
by violating the spatial translational invariance and 
the total momentum conservation, 
which also change the number of zero-Lyapunov exponents 
\cite{Tan03a}.  
   As a second point, some mode structures were observed in 
some of the above Lyapunov vector components, which 
should be constant in zero-Lyapunov exponents. 
   For example, a mode structure in 
the Lyapunov vector component $\delta y_{j}^{(n)}$
(the transverse spatial translational invariance 
Lyapunov mode) is well known \cite{Pos00}. 
   This mode is stationary in time, and appears in one 
of the two types of the Lyapunov steps. 
   Ref. \cite{Tan03a} showed another mode structure 
in $\delta y_{j}^{(n)}/p_{yj}$, 
(the transverse time translational Lyapunov mode).   
   This mode depends on time, and appears 
in other types of the Lyapunov steps. 
   These Lyapunov modes are enough to categorize 
all the Lyapunov steps.  
   Ref. \cite{Pos00,For04} also claim a moving mode structure 
in $\delta x_{j}^{(n)}$. 

   However, there has not been enough evidence yet to confirm
the above scenario for the Lyapunov steps and modes. 
   For example, the mode structure in 
the momentum part of Lyapunov vectors 
has not been reported explicitly. 
   Besides, the phase relations of different modes, 
for example, the modes in $\delta y_{j}^{(n)}/p_{yj}$ 
and $\delta x_{j}^{(n)}$, have not been discussed. 
   Another important problem is 
the time scale specified by a time-dependent  
Lyapunov mode, like the time-oscillation for the 
mode in $\delta y_{j}^{(n)}/p_{yj}$. 
   The time-oscillating period is usually 
much longer than the mean free time of the system, and 
it should correspond to 
a collective movement, 
but quantitative evidence for it 
has not been shown explicitly.  
   
   As an indicator for collective movements of many-particle 
systems, we can use the momentum auto-correlation functions, 
where collective movement may appear as a time-oscillation 
behavior \cite{Zwa67}, as observed in many macroscopic models 
\cite{Rah67,Lev73,Ber81,All83,Han90}. 
   The auto-correlation 
functions are accessible experimentally using 
neutron and light scattering techniques 
\cite{Han90,Cop75,Wel85,Sch99}. 
   As an essential aspect of auto-correlation functions is 
their role as response functions for the system. 
   For example, linear response theory connects the 
time integral of the auto-correlation function  
with a transport coefficient  \cite{Kub78}. 
   However it should be emphasized that the auto-correlation 
functions themselves provide much more detailed information 
about the system. 
   Linear response theory requires the time-integral 
and the thermodynamic limit of the auto-correlation function 
to calculate transport coefficients, and in this process 
information about short time-scales and a finite size effects 
are lost. 
   For instance, the time-oscillation of the auto-correlation 
functions is one of the finite size 
effects, which linear response theory does not treat. 
   Information on short time-scales and finite size effects 
in the auto-correlation functions also plays an important role 
in the generalized hydrodynamics \cite{Sch88} and 
generalized Fokker-Planck equation \cite{Zwa61}. 
   
   The purposes of this paper are two-fold.  
   First we calculate all components 
of the Lyapunov vectors associated 
with the stepwise structure of the Lyapunov spectrum. 
   They include the longitudinal and transverse components 
of both the spatial and momentum parts of the Lyapunov vectors. 
   We demonstrate the wavelike structures in the 
components of the Lyapunov vectors, 
and specify their phase relations.  
   These results support the above explanation for 
the origin of the Lyapunov steps and 
modes based on the zero-Lyapunov exponents and the 
associated Lyapunov vectors. 
   Spatial node structures of these Lyapunov modes 
are explained in terms of boundary conditions. 
   It is emphasized that some of the Lyapunov modes 
show time-oscillating behaviors, and a particle number 
dependence in their time-oscillating periods. 
   The second purpose of this paper is to 
discuss the connection between the time-oscillation 
of the Lyapunov modes and the momentum auto-correlation 
functions. 
   Our central result is that the longest 
time-oscillating period of the Lyapunov modes is 
twice as long as that for the auto-correlation function 
for the longitudinal component of the momentum. 
   This relation is satisfied irrespective of 
the number of particles in the system and the boundary conditions. 
   We give a simple explanation for this relation. 
   This result means that the time-oscillating 
behavior of the Lyapunov vectors is reflected 
by a collective movement within many-particle systems. 
   This also gives some evidence to connect the Lyapunov mode, 
a tangent space property specified by Lyapunov vectors, 
to the auto-correlation function, 
which is a phase space property and is accessible 
experimentally. 

   We use a quasi-one-dimensional many-hard-disk system, 
as this model allows fast numerical calculation of the Lyapunov 
exponents and vectors, and shows clear 
Lyapunov steps and modes. 
   In general the numerical calculation of the Lyapunov spectrum 
and vectors are very time-consuming, and it is often  
difficult to get clear Lyapunov mode structures, 
particularly for time-dependent Lyapunov modes. 
   If the above picture of the Lyapunov steps 
and modes based on universal properties 
such as the translational invariances and the conservation 
laws can be justified, then a simple model should be
sufficient to convince us of their origin. 
   This system also exhibits a clear oscillatory 
behavior in the longitudinal momentum auto-correlation 
function.  
   A useful technique to get clear Lyapunov steps 
and modes is to use hard-wall boundary conditions. 
   Although hard-wall boundary conditions destroy the  
spatial translational invariance and the total momentum 
conservation, and lead to different structure 
in the Lyapunov spectrum compared to periodic boundary 
conditions, it has been shown that there is a simple  
relation between the observed Lyapunov steps and modes 
and different boundary conditions \cite{Tan03a}. 
   Specifically, we use a quasi-one-dimensional system 
with hard-wall boundary conditions in the longitudinal direction 
and periodic boundary conditions in the transverse direction.  
   Usually, the hard-wall boundary conditions make numerical 
calculation slower compared to periodic boundary conditions, 
but in our system only the two particles at each end of the system 
collide with the hard-walls so the effect is small. 

   The outline of this paper is as follows.  
   In Sec. \ref{QuasiOneDimSystem}, the quasi-one-dimensional 
system is introduced. 
   In Sec. \ref{LyapunovStepsModes} we discuss the 
Lyapunov steps and modes. 
   In Sec. \ref{AutoCorrelationFunctions} 
the momentum auto-correlation functions and their 
relation with the Lyapunov modes is discussed . 
   Finally we give some conclusion and remarks 
in Sec. \ref{ConclusionRemarks}.

%
%
\section{Quasi-one-dimensional system}
\label{QuasiOneDimSystem}


   The model considered in this paper is a quasi-one-dimensional 
many-hard-disk system. 
   It is a two-dimensional rectangular system consisting 
of many hard-disks with the width of the system so narrow 
that the disk positions are not exchanged, thus the disks
can be numbered from left to right. 
   We assume that the mass $m$ and the radius $R$ of each disk
is the same. 
   In this case the quasi-one-dimensional system has a width $L_{y}$ 
that satisfies the condition $2R < L_{y} < 4R$.
   Figure \ref{figA1qua1dim} gives a schematic illustration 
of the quasi-one-dimensional system with the particles numbered  
$1,2,\cdots,N$ ($N$: particle number) from 
the left to right. 

\begin{figure}[!htb]
\vspfigA
\caption{
      A schematic illustration of a quasi-one-dimensional system 
used in this paper.  
      The system shape is so narrow that particles always remain 
   in the same order. 
      Here, $L_{x}$ ($L_{y}$) is the length (width) of the system 
   in the longitudinal (transverse) direction, 
   and $R$ is the radius 
   of a particle. 
      The dashed lines represent periodic boundary conditions
 and the solid lines represent hard-wall boundary conditions. 
      The particles are numbered $1,2,\cdots,N$ 
   from the left to the right. 
   }
\vspfigB
\label{figA1qua1dim}\end{figure}  


   Originally, the quasi-one-dimensional many-hard-disk system 
was introduced as a system to easily and clearly observe 
the stepwise structure of Lyapunov spectrum 
(which we call the "Lyapunov steps") and the 
corresponding wave-like structure of Lyapunov vectors 
(the so called "Lyapunov modes") \cite{Tan03a}. 
   Numerical observation of the Lyapunov steps and modes is 
very time-consuming, and even at present the Lyapunov spectra 
is limited to about 1000 particles \cite{Pos00}. 
   Therefore it is valuable to explore fast and 
effective ways to calculate them numerically. 
   It is well known that the rectangular system has a wider 
stepwise region than a square system with the same area 
and number of particles. 
This quasi-one-dimensional system is the most 
rectangular two-dimensional system possible. 
   Ref.  \cite{Tan03a} demonstrated that 
in the quasi-one-dimensional system we can clearly observe 
the structure of the Lyapunov modes. 
   It is known that there are two kinds 
of Lyapunov modes: stationary modes and time-dependent modes. 
   The stational Lyapunov modes are most easily observed 
because of their stable structure, but 
generally the observation of the time-dependent Lyapunov modes 
is much harder because of large fluctuations in their 
structure and their intrinsic time-dependence. 
   The quasi-one-dimensional system 
is the first system in which time-oscillating Lyapunov modes 
were demonstrated \cite{Tan03a}. 

   Another advantage of the quasi-one-dimensional system is that  
the particle interactions in this system are restricted  
to nearest-neighbor particles only, so we need much less 
effort to find colliding particle pairs numerically 
compared with the fully two-dimensional system in which 
each particle can collide with any other particle. 
   This leads to faster numerical calculation of the 
system dynamics. 
   Besides, in this system, particle movement in 
the narrow direction is suppressed, compared to 
the longitudinal  direction, and roughly speaking the particle sequence  
corresponds to the particle position. 
   This leads to a much simpler representation of Lyapunov modes, 
which must be investigated as functions of spatial coordinates 
and time.  
   One may also notice that in the quasi-one-dimensional system 
the system size is proportional to the particle number $N$, 
while for the square system it is proportional to $\sqrt{N}$. 
   This implies that in the quasi-one-dimensional system, 
many-particle effects more evident than for the fully 
two-dimensional systems with the same number of particles.


   Another important point in the quasi-one-dimensional system 
is the effect of boundary conditions. 
   Different from the square system, in which the boundary length 
is proportional to the square root of the system size, 
in the quasi-one-dimensional system the boundary length is 
proportional to the system size itself, therefore 
we cannot neglect its effect even in the thermodynamic limit. 
   Actually Ref. \cite{Tan03a} showed that the Lyapunov steps 
and modes depend strongly on boundary conditions. 
   Boundary conditions change, not only the structure of 
Lyapunov steps and modes, but also the clearness of Lyapunov mode 
structure. 
   For example, a system with purely hard-wall boundary conditions 
has no stationary Lyapunov mode and its corresponding Lyapunov 
steps and show much clearer time-oscillating Lyapunov modes,  
compared to a system with the periodic boundary conditions. 
   Roughly speaking, hard-wall boundary conditions pin the 
positions of the nodes and thus lead to clearer Lyapunov modes. 
   On the other hand, the numerical calculation with the hard-wall 
boundary conditions is more time-consuming. 
   This disadvantage is significant in the quasi-one-dimensional 
system with purely hard-wall boundary conditions.
   As another disadvantage of the system with purely 
hard-wall boundary conditions, we cannot investigate the stationary 
Lyapunov mode due to spatial translational invariance. 
   As an optimal system we mostly consider a quasi-one-dimensional 
system with hard-wall boundary 
conditions in the longitudinal direction and 
periodic boundary conditions in the transverse direction.  
   We use the notation (H,P) for this boundary condition 
throughout this paper. 
   In Fig. \ref{figA1qua1dim} we represent  the boundary condition (H,P) 
as different types of lines on the boundaries: 
the bold solid lines signify hard-wall boundary conditions 
and the broken lines signify periodic boundary conditions. 
   In this system we can get much clearer Lyapunov modes 
than for purely periodic boundary conditions and we observe 
both types of Lyapunov steps and modes.


   Although the quasi-one-dimensional many-hard-disk system 
with the boundary condition (H,P) may be artificially introduced 
to investigate Lyapunov steps and modes in a fast and effective way, 
it is essential to note that the results from this system can be used 
to predict  Lyapunov steps and modes in more general systems, 
such as a fully two-dimensional system with purely periodic boundary 
conditions. 
   Details of the relation of the Lyapunov steps and modes in 
quasi-one-dimensional systems with different boundary conditions 
were given in Ref. \cite{Tan03a}. 
   For example, the step widths of the Lyapunov spectrum (the spatial 
and time periods of the corresponding Lyapunov modes) in the system 
with the boundary condition (H,P) are halves (twice) the ones 
in the system with the purely periodic boundary conditions (P,P). 
   It is also known that the structure of Lyapunov steps 
for the quasi-one-dimensional system is the same as the fully 
two-dimensional system. 

   As we discussed above, the main reason to use the 
quasi-one-dimensional system with the boundary condition (H,P) 
is to get clear Lyapunov steps and modes 
in a fast numerical calculation.  
   On the other hand, these advantages may not assist the  
calculation of the momentum auto-correlation function.  
   In the quasi-one-dimensional system, collisions of a particle 
are restricted to its two nearest-neighbor particles only, 
so it may be supposed that specific types of collisions 
like the "back scattering effect" play an important role 
in this system. 
   The back scattering effect, which comes 
from a reversal of the velocity of a particle 
by a collision with the nearest neighbor particle, 
can lead to a negative region of the momentum auto-correlation 
function.     
   One may also suppose that in the quasi-one-dimensional system 
a collective motion may be enhanced, because the movement 
of particles in the narrow direction is very restricted. 
   This may lead to a clear time-oscillation of the momentum 
auto-correlation function, as will be actually 
observed in Sect. \ref{AutoCorrelationFunctions} of this paper.  
   It may also be noted that boundary condition effects on the 
momentum auto-correlation function are not well known.  
   Because we need to know about them to be able to 
guess the relation between the auto-correlation functions 
and Lyapunov modes in different boundary conditions,  
we  will discuss boundary condition effects 
on momentum auto-correlation function briefly in 
Sect. \ref{BoundaryConditionEffects}. 

   In this paper we use units where the mass $m$ and 
the particle radius $R$ are $1$, and the total energy $E$ is $N$ 
(except in Sec. \ref{LyapunovModeAmplitude}). 
   For the numerical calculations, the system lengths are 
chosen as $L_{x} = 1.5 NL_{y} + 2R$ and $L_{y} = 2R(1+10^{-6})$ 
for the quasi-one-dimensional system 
with the boundary condition (H,P).

\section{Lyapunov Steps and Modes} 
\label{LyapunovStepsModes}

   In this section we discuss the Lyapunov steps and modes in 
the quasi-one-dimensional system with boundary condition (H,P). 
   Part of these results have already 
been presented in Ref. \cite{Tan03a}, 
and here we complete this presentation.  
   Some of the discussions omitted in Ref. \cite{Tan03a} were 
the relation between the time-oscillating Lyapunov mode 
proportional to the momentum and the longitudinal Lyapunov modes, 
the Lyapunov modes for the momentum parts of Lyapunov vectors, 
and the particle number dependence of the time-oscillating period 
of the Lyapunov modes.  

   We introduce the Lyapunov spectrum as the ordered set of the Lyapunov
exponents $\lambda^{(n)}$, $n=1,2,\cdots,4N$, where
$\lambda^{(1)} \geq \lambda^{(2)} \geq \cdots \geq \lambda^{(4N)}$. 
   The notations $\delta \mathbf{q}_{j}^{(n)} 
\equiv ( \delta x_{j}^{(n)}, \delta y_{j}^{(n)} )$ 
and $\delta \mathbf{p}_{j}^{(n)} 
\equiv ( \delta p_{xj}^{(n)}, \delta p_{yj}^{(n)} )$
are used for the spatial and 
momentum components, respectively,  of  the  $j$-th particle Lyapunov vector 
corresponding to the $n$-th Lyapunov exponent $\lambda^{(n)}$. 

   For a numerical calculation of the Lyapunov spectrum and 
the Lyapunov vectors, we used the numerical algorithm developed by 
Benettin et al. \cite{Ben76} and Shimada et al. \cite{Shi79} 
(Also See Refs. \cite{Ben80a,Ben80b}).
   This algorithm is characterized by regular 
re-orthogonalizations and renormalization of set of Lyapunov 
vectors, which can be done after each particle collision in 
a many-hard-disk system. 
   Usually the Lyapunov steps and modes appear after 
a long trajectory calculation, and we typically calculated 
trajectories of more than $5\times 10^{5}$ particle 
collisions to get the Lyapunov spectra and vectors 
shown here. 

   The main purpose in this section is to investigate 
the time-oscillating structures in the transverse and 
longitudinal Lyapunov modes. 
   As we mentioned in Sect. \ref{QuasiOneDimSystem} 
the numerical calculation of Lyapunov spectra and 
Lyapunov vectors for many-particle systems is 
time-consuming even in the quasi-one-dimensional system, 
and in this paper we present results of $100$ particle systems. 
   In such a rather small system the structure of the Lyapunov 
modes has large fluctuations in space and time, 
which prevents the appearance of clear mode structures. 
   Another problem appears in the investigation of 
the Lyapunov modes proportional to the momentum. 
   We investigate these modes by calculating 
such the quantities $\delta x_{j}^{(n)}/p_{xj}$, which 
should show a time-oscillating wave-like structure, 
if the Lyapunov vector components has modes 
proportional to momentum. 
   However, fluctuations of this quantity 
may be enhanced in cases where the momentum component
has a small absolute value.  
   To overcome these technical problems, and 
to visualize this structure as clearly as possible, 
we take a local time-average of the Lyapunov modes. 
   More concretely, for investigations of the modes 
of the Lyapunov vector components $\delta x_{j}^{(n)}$ 
and $\delta y_{j}^{(n)}$ we take their arithmetic average 
over $8N$ collisions using data just after collisions, and 
plot them as functions of the same local time 
arithmetic average of $x_{j}$ and the first collision number of 
time interval taking the local time average. 
   For modes in the quantities $\delta x_{j}^{(n)}/p_{xj}$ 
$\delta y_{j}^{(n)}/p_{yj}$, $\delta p_{yj}^{(n)}/p_{xj}$, 
and $\delta p_{yj}^{(n)}/p_{yj}$ we take the same local time 
average, except that if the absolute value $|p_{xj}|$ 
($|p_{yj}|$) of the momentum component is less than $10\%$ 
of the averaged momentum amplitude $\sqrt{2ME/N}$ then we 
exclude the data at that time from the local 
time average. 
   We use the notation $\langle \cdots \rangle_{t}$ for 
such a local time average in this paper.

\subsection{Lyapunov steps}

   Figure \ref{lyapu} is the stepwise structure of the Lyapunov spectrum 
normalized by the largest Lyapunov exponent 
for the quasi-one-dimensional many-hard-disk system of $100$ particles 
with (H,P) boundary condition. 
   The entire positive branch of the Lyapunov spectrum 
is shown in the inset to this figure. 
   In a Hamiltonian system, the negative branch of the Lyapunov 
spectrum takes the same absolute value as the positive branch of the 
Lyapunov spectrum from the conjugate pairing rule: 
$\lambda^{(4N-n+1)} = - \lambda^{(n)}$, $n = 1,2,\cdots,2N$ 
\cite{Arn89}, so they are omitted in Fig. \ref{lyapu}. 

\begin{figure}[!htb]
\vspfigA
\caption{
      Stepwise structure of the Lyapunov spectrum 
   normalized by the largest Lyapunov exponent 
   for the quasi-one-dimensional system 
   with hard-wall boundary conditions in $x$-direction 
   and periodic boundary conditions in $y$-direction. 
      The particle number is $N=100$.
      The white-filled (black-filled) circles correspond 
   to the stational (time-oscillating) Lyapunov modes. 
      Inset: The full spectrum of the positive branch of 
   the normalized Lyapunov spectrum. 
   }
\vspfigB
\label{lyapu}\end{figure}  

   This system has 4 zero-Lyapunov exponents, which come 
from the conservation of the $y$-component of the total momentum 
and the center of mass, energy conservation, and the 
deterministic nature of the orbit. 
   Note that  the $x$-component of the total momentum and the center 
of mass are not conserved, because of the hard-wall boundary 
condition in the $x$-direction. 
   Half of these 4 zero-Lyapunov exponents appear 
in Fig. \ref{lyapu}.  
   
   The stepwise structure of the Lyapunov spectrum in this system 
is one and two-point steps. 
   These two kinds of Lyapunov steps accompany different 
mode structures in the Lyapunov vectors: 
one is the stationary modes, as discussed in subsections 
\ref{StationaryLyapunovModes}, and the other is time-oscillating 
modes, as discussed in the subsection 
\ref{TimeOscillatingLyapunovModes}. 
   Here we count the sequence of Lyapunov steps 
from the zero Lyapunov exponents, so $\lambda^{(200)}$ and
$\lambda^{(199)}$ are the zero exponents,
$\lambda^{(198)}$ is the first one-point step, 
$\lambda^{(197)}$ and $\lambda^{(196)}$  are the first two-point step,
$\lambda^{(195)}$ is the second one-point step, and
$\lambda^{(194)}$ and $\lambda^{(193)}$ are the second two-point step,
see Fig. \ref{lyapu}. 
%
%
\subsection{Stationary Lyapunov modes}
\label{StationaryLyapunovModes}

   First we discuss the Lyapunov mode corresponding to the 
first and second one-point steps in Fig. \ref{lyapu}. 
Fig. \ref{figB2modestat} shows the graph of the 
Lyapunov vector components corresponding to the first 
and second one-point step 
as a function of the collision number 
$n_{t}$ and the normalized local time average 
$\langle x_{j} \rangle_{t}/L_{x}$ of the $x$-component 
of the particle position. 
   Here, Fig. \ref{figB2modestat}(a) is 
for the local time-averages of 
$\langle \delta x_{j}^{(198)} \rangle_{t}$ and 
$\langle \delta y_{j}^{(198)} \rangle_{t}$ for 
the first one-point step $\lambda^{(198)}$, and 
Fig. \ref{figB2modestat}(b) is a similar graph 
for the second one-point step $\lambda^{(195)}$. 
   These one-point steps are indicated by arrows in Fig. 
\ref{lyapu}.
   Both graphs have the same collision number interval 
$[524000,569600]$. 
   On the base of each of Figs. \ref{figB2modestat}(a) and (b) 
we give contour-plots of the 
transversal modes 
$\langle \delta y_{j}^{(198)} \rangle_{t}$ and 
$\langle \delta y_{j}^{(195)} \rangle_{t}$, respectively,  
in which the dotted lines, 
the solid lines, and the broken lines correspond to 
the levels -0.08, 0, and +0.08, respectively.  

\begin{figure}[!htb]
\vspfigA
\caption{
      Local time averages of $\langle \delta x_{j}\rangle_{t}$ 
   and $\langle \delta y_{j} \rangle_{t}$ for the first two-point
   step (exponent 198) in (a), 
   and for the second two-point step (exponent 195) in (b). 
  These are shown as functions of the collision number $n_{t}$ and the 
   normalized local time average $\langle x_{j} \rangle_{t} 
   /L_{x}$ of the $x$-component of the $j$-th particles position
   (with the system length $L_{x}$). 
      The corresponding Lyapunov exponents $\lambda^{(198)}$ 
   and $\lambda^{(195)}$ are indicated by arrows in Fig. 
   \ref{lyapu}. 
      The base of each graph are the contour plots of the 
   transverse Lyapunov modes at the levels $-0.08$ (broken lines), 
   $0$ (solid lines), and $+0.08$ (dotted lines). 
   }
\vspfigB
\label{figB2modestat}\end{figure}  

   In Fig. \ref{figB2modestat} we recognize spatial wave-like 
structures in the transverse components of  
$\langle \delta y_{j}^{(n)} \rangle_{t}$ for 
$n=198$ and $195$ which are stationary in time 
(at least in a time interval of more than $45\times 10^{3}$ 
collisions as shown in this figure). 
   These wave-like structures are very nicely fitted 
by sinusoidal functions \cite{Tan03a}. 
   Note that in the calculation of Lyapunov vectors 
we used the numerical algorithm which imposed  
renormalization of the Lyapunov vectors at every collision, 
so that the amplitudes of any component of the Lyapunov vector must 
be less than 1. 
   It should be emphasized that anti-nodes in the modes appear 
at the end of the system in the $x$-direction.  
   By comparison the amplitudes of the longitudinal components of 
$\langle \delta x_{j}^{(n)} \rangle_{t}$ for $n=198$ 
and $195$ are extremely small. 
   These observations suggest that the Lyapunov mode 
corresponding to the $k$-th one-point Lyapunov step is 
approximately represented by  

\begin{eqnarray}
   \delta x_{j}^{(\mu (k))}  &\approx& 0,   
   \label{LyapuWaveWPx} \\ 
   \delta y_{j}^{(\mu (k))}  &\approx& 
   \alpha_{k} \cos\left(\frac{\pi k}{L_{x}} x_{j} \right) ,   
\label{LyapuWaveWPy}\end{eqnarray}

\noindent $j=1,2,\cdots,N$,
corresponding to the Lyapunov exponents $\lambda^{(\mu(k))}$, 
in which $\alpha_{k}$ is a constant and $\mu(k)$ is 
the Lyapunov index corresponding to the $k$-th 
one-point step of the Lyapunov spectrum. 
   Here we take the origin of the $x$-component of 
the spatial coordinate to be $x_{j}=0$, so that 
an ambiguity in spatial phase can be removed in 
Eq. (\ref{LyapuWaveWPy}).

\subsection{Time-oscillating Lyapunov modes}
\label{TimeOscillatingLyapunovModes}

   Now we discuss the remaining Lyapunov modes, which correspond 
to the two-point steps of the Lyapunov spectrum. 
   Figure \ref{figB3modeoscil3d} shows the graphs of the local 
time averages of 
$\langle \delta x_{j}^{(197)} \rangle_{t}$, 
$\langle \delta x_{j}^{(197)}/p_{xj} \rangle_{t}$, 
$\langle \delta y_{j}^{(197)} \rangle_{t}$, and
$\langle \delta y_{j}^{(197)}/p_{yj} \rangle_{t}$ 
as functions of the collision number $n_{t}$ and normalized 
local time averaged position $\langle x_{j} \rangle_{t}/L_{x}$ 
of the $j$-th particle. 
   The number of particles is $N=100$, and the four graphs are 
for the same collision number interval $n_{t} \in [535200,569600]$. 
   These correspond to the first exponent on the first  two-point
step, exponent $197$, 
in Fig. \ref{lyapu}. 
   In these figures, we draw a contour plot on the base  
of each three dimensional 
graph at levels $-0.08$ (dotted lines), $0$ (solid lines), 
and $+0.08$ (broken lines).

\begin{figure}[!htb]
\vspfigA
\caption{
      Local time averages of 
   $\langle \delta x_{j} \rangle_{t}$, 
   $\langle \delta x_{j}/p_{xj} \rangle_{t}$, 
   $\langle \delta y_{j} \rangle_{t}$, and
   $\langle \delta y_{j}/p_{yj} \rangle_{t}$
   for the Lyapunov modes corresponding to the 
   first exponent of the first two-point step
   ($\lambda^{(197)}$), 
   as functions of the collision number $n_{t}$ and 
   the normalized local time average 
   $\langle x_{j} \rangle_{t}/L_{x}$ of 
   the $x$-component of the position of the $j$-th 
   particle. 
      The base of each graph is a contour plot of the 
   three dimensional graph at the levels $-0.08$ (dotted lines), 
   $0$ (solid lines), and $+0.08$ (broken lines). 
   }
\vspfigB
\label{figB3modeoscil3d}\end{figure}  

   We can easily recognize spatial wave-like structures with 
time-oscillations in Figs. \ref{figB3modeoscil3d}(a) and (d). 
   In Fig. \ref{figB3modeoscil3d}(a), the 
longitudinal Lyapunov vector component 
$\langle \delta x_{j}^{(197)} \rangle_{t}$ has nodes at the ends  
of the quasi-one-dimensional system, and 
the wave-length is given by $2L_{x}$. 
   On the other hand, in Fig. \ref{figB3modeoscil3d}(d), 
the transverse Lyapunov vector component 
$\langle \delta y_{j}^{(197)}/p_{yj} \rangle_{t}$ 
has anti-nodes at the end of 
the system and has a node at the middle, 
although its wave-length is given by $2L_{x}$ as for  
the longitudinal mode of Fig. \ref{figB3modeoscil3d}(a). 

   There is also a time-oscillating wave-like structure 
in the longitudinal Lyapunov vector component 
$\langle \delta x_{j}^{(197)}/p_{xj} \rangle_{t}$, as shown 
in Fig. \ref{figB3modeoscil3d}(b). 
   This structure is different from the one shown 
in Fig. \ref{figB3modeoscil3d}(a) associated with the same 
longitudinal Lyapunov vector component $\delta x_{j}^{(197)}$. 
   It has anti-nodes at the ends of the system, and has a node in the middle. 
   Its wave-length is $2 L_{x}$. 
   These characteristics suggest that although there are 
large fluctuations in the middle of the system, 
the time-oscillating wave-like structure 
in Fig. \ref{figB3modeoscil3d}(b) is the same as that in 
Fig. \ref{figB3modeoscil3d}(d). 
   In Fig. \ref{figB3modeoscil3d}(c) it is rather difficult to 
recognize any structure. 
   Roughly speaking, it is just random fluctuations, 
in middle of the system but the 
amplitude of such fluctuations is small compared to 
region at the end of the system. 
   However such small amplitude fluctuations are  
required for consistency with Fig. \ref{figB3modeoscil3d}(d), 
namely the fact that in this region 
the value of $\delta y_{j}^{(197)}/p_{yj}$ is small, 
so the value of $\delta y_{j}^{(197)}$ itself should be small 
with an almost position independent momentum $p_{yj}$.  


   Next we discuss the phase relations for the Lyapunov modes of
the first two-point steps, corresponding to exponents 
$\lambda^{(196)}$ and $\lambda^{(197)}$
(the black-filled circles with brace underneath
in Fig. \ref{lyapu}). 
   Figure \ref{figB4modeoscil2dconto} shows the contour plots 
of the local time averages of 
$\langle \delta x_{j} \rangle_{t}$, 
$\langle \delta x_{j}/p_{xj} \rangle_{t}$, 
$\langle \delta y_{j}/p_{yj} \rangle_{t}$, and
for $\lambda^{(196)}$ and $\lambda^{(197)}$
as functions of the collision number $n_{t}$ and the normalized 
local time average $\langle x_{j} \rangle_{t}/L_{x}$ of 
the position of the $j$-th 
particle at levels $-0.08$, $0$, and $+0.08$, for a  
$100$-particle system. 
   The six graphs in Fig. \ref{figB4modeoscil2dconto} have 
the same collision number interval $[535200,569600]$, 
and Figs. \ref{figB4modeoscil2dconto}(a), (c), and (e) 
correspond to Figs. \ref{figB3modeoscil3d}(a), (b), and (d), 
respectively. 

\begin{figure}[!htb]
\vspfigA
\caption{ 
      Contour plots of the local time averages of
   $\langle \delta x_{j} \rangle_{t}$, 
   $\langle \delta x_{j}/p_{xj} \rangle_{t}$, 
   $\langle \delta y_{j}/p_{yj} \rangle_{t}$, 
   for the first two-point steps ($\lambda^{(197)}$ (a,c,e) 
   and $\lambda^{(196)}$ (b,d,f)), 
   as functions of the collision number $n_{t}$ and 
   the normalized local time average 
   $\langle x_{j} \rangle_{t}/L_{x}$ of 
   the position of the $j$-th 
   particle in the same collision number interval 
   $[535200,569600]$. 
      The dotted lines, the solid lines, and the broken lines 
   are contour lines at the levels $-0.08$, $0$, and $+0.08$, 
   respectively. 
   }
\vspfigB
\label{figB4modeoscil2dconto}\end{figure}  

   Figure \ref{figB4modeoscil2dconto} shows that the two Lyapunov 
exponents for the same 
two-point step have the same structure of Lyapunov modes, 
but they are orthogonal in time, namely node lines of 
the Lyapunov modes corresponding to the Lyapunov exponent 
$\lambda^{(197)}$ correspond to anti-node lines of the 
ones of $\lambda^{(196)}$. 
   We also notice that the node lines of the 
Lyapunov modes in 
$\langle \delta x_{j}^{(n)}/p_{xj} \rangle_{t}$ and 
$\langle \delta y_{j}^{(n)}/p_{yj} \rangle_{t}$ 
coincide with each other in space and time ($n=197,196$), 
on the other hand the Lyapunov modes in 
$\langle \delta x_{j}^{(n)}\rangle_{t}$ and 
$\langle \delta x_{j}^{(n)}/p_{xj} \rangle_{t}$ 
are orthogonal in space and time at the same Lyapunov index. 

   The above discussions based on Figs. \ref{figB3modeoscil3d} 
and \ref{figB4modeoscil2dconto} (and similar observations 
of Lyapunov modes in the other two-point steps of the 
Lyapunov spectrum) lead to the conjecture that the 
spatial part of Lyapunov vector 
components $\delta x_{j}^{(\nu (k))}$ and 
$\delta y_{j}^{(\nu (k)-1)}$ corresponding to the 
Lyapunov exponents constructing the $k$-th two-point step 
are approximately expressed as 

\begin{widetext}
\begin{eqnarray}
   \delta x_{j}^{(\nu (k))} &\approx& 
   \alpha_{k}' \; p_{xj}
	     \cos\left(\frac{\pi k}{L_{x}} x_{j} \right) 
	     \cos\left( \frac{2\pi k}{T_{lya}} n_{t}
            \!+\! \beta_{k}' \right) 
   + \tilde{\alpha}_{k}' \; 
	     \sin\left(\frac{\pi k}{L_{x}} x_{j} \right) 
	     \sin\left( \frac{2\pi k}{T_{lya}} n_{t}
            \!+\! \beta_{k}' \right) 
\label{LyapuOscilWPx1} 
\\
   \delta x_{j}^{(\nu (k)-1)} &\approx& 
   \alpha_{k}'' \; p_{xj}
	     \cos\left(\frac{\pi k}{L_{x}} x_{j} \right) 
	     \sin\left( \frac{2\pi k}{T_{lya}} n_{t}
            \!+\! \beta_{k}' \right) 
   + \tilde{\alpha}_{k}'' \; 
	     \sin\left(\frac{\pi k}{L_{x}} x_{j} \right) 
	     \cos\left( \frac{2\pi k}{T_{lya}} n_{t}
     \!+\! \beta_{k}' \right) 
\label{LyapuOscilWPx2} 
\end{eqnarray}
\begin{eqnarray}
   \delta y_{j}^{(\nu (k))} &\approx& 
   \alpha_{k}' \; p_{yj}
	     \cos\left(\frac{\pi k}{L_{x}} x_{j} \right) 
	     \cos\left( \frac{2\pi k}{T_{lya}} n_{t}
            \!+\! \beta_{k}' \right) 
\label{LyapuOscilWPy1} \\
   \delta y_{j}^{(\nu (k)-1)} &\approx& 
   \alpha_{k}'' \; p_{yj}
	     \cos\left(\frac{\pi k}{L_{x}} x_{j} \right) 
	     \sin\left( \frac{2\pi k}{T_{lya}} n_{t}
            \!+\! \beta_{k}' \right)
\label{LyapuOscilWPy2}
\end{eqnarray}
\end{widetext}

\noindent $j=1,2,\cdots,N$ with constants $\alpha_{k}'$, 
$\alpha_{k}''$, $\tilde{\alpha}_{k}'$, $\tilde{\alpha}_{k}''$ 
and $\beta_{k}'$. 
   It should be noted that large fluctuations in the 
Lyapunov mode represented in middle of 
Figs \ref{figB3modeoscil3d}(b), \ref{figB4modeoscil2dconto}(c), 
and \ref{figB4modeoscil2dconto}(d) can come from the second terms 
on the right-hand sides of Eqs. (\ref{LyapuOscilWPx1}) and 
(\ref{LyapuOscilWPx2}).  
   On the other hand the effect of the first terms on the 
right-hand side of Eqs. (\ref{LyapuOscilWPx1}) and 
(\ref{LyapuOscilWPx2}) does not appear explicitly in 
Figs \ref{figB3modeoscil3d}(a), \ref{figB4modeoscil2dconto}(a) 
and \ref{figB4modeoscil2dconto}(b), because 
the factor $p_{xj}$ in these terms distributes their contributions 
randomly and these terms disappear after taking 
local time averages. 

\subsection{Energy dependence of Lyapunov mode amplitudes}
\label{LyapunovModeAmplitude}

   In the expressions (\ref{LyapuOscilWPx1}), 
(\ref{LyapuOscilWPx2}), (\ref{LyapuOscilWPy1}) 
and (\ref{LyapuOscilWPy2}) for the time-oscillating 
Lyapunov modes, the quantities $\alpha'_{k}$, $\alpha''_{k}$, 
$\tilde{\alpha}'_{k}$ and $\tilde{\alpha}''_{k}$ are 
introduced simply as coefficients of the linear combination of the 
longitudinal spatial translational invariance Lyapunov mode 
and the time translational invariance Lyapunov mode.  
   However it is important to note that these coefficients are 
related to each other through the normalization 
of the Lyapunov mode. 

   We consider the condition that the Lyapunov mode vector 
$(\delta x_{1}^{(\nu')},$ $\delta x_{2}^{(\nu')},
\cdots,\delta x_{N}^{(\nu')}, $ $
\delta y_{1}^{(\nu')},$ $\delta y_{2}^{(\nu')},$ $ 
\cdots,\delta y_{N}^{(\nu')})$, $\nu' = \nu (k), \nu (k)-1$ 
is normalizable. 
   This leads to the approximate relations  
   
\begin{eqnarray}
   |\alpha'_{k}| &\sim& 
   \frac{|\tilde{\alpha}'_{k}|}{\sqrt{2mE/N}}, 
   \label{normaLyavec1}\\
   |\alpha'_{k}| &\sim& |\alpha''_{k}|, 
   \label{normaLyavec2}\\
   |\tilde{\alpha}'_{k}| &\sim& |\tilde{\alpha}''_{k}|
\label{normaLyavec3}\end{eqnarray}
 
\noindent with the mass $m(=1)$, the total energy $E$ 
and the number of particles $N$.

 \begin{figure}[!htb]
\vspfigA
\caption{ 
      Amplitudes $|\alpha'_{1}|$ and $|\tilde{\alpha}'_{1}|$ 
   of Lyapunov modes  
   $\langle \delta x_{j}^{(2N-3)} \rangle_{t}$        (circles), 
   $\langle \delta x_{j}^{(2N-3)}/p_{xj} \rangle_{t}$ (triangles), 
   and $\langle \delta y_{j}^{(2N-3)}/p_{yj} \rangle_{t}$ (squares) 
   as functions of $\sqrt{2mE/N}$  
   in the quasi-one-dimensional system of $50$ hard-disks 
   with (H,P) boundary condition. 
      The broken line is a fit of the amplitude 
   $|\tilde{\alpha}'_{1}|$ to a constant function $y=\xi$ 
   with a fitting parameter $\xi$, and the solid line is 
   given by $y=\xi x$.   
      }
\vspfigB
\label{figB5oscillampli}\end{figure}  

   In Fig. \ref{figB5oscillampli} we show the amplitudes 
$|\alpha'_{1}|$ and $|\tilde{\alpha}'_{1}|$, which are obtained 
by fitting the Lyapunov modes 
$\langle \delta x_{j}^{(2N-3)} \rangle_{t}$, 
$\langle \delta x_{j}^{(2N-3)}/p_{xj} \rangle_{t}$, 
and $\langle \delta y_{j}^{(2N-3)}/p_{yj} \rangle_{t}$ 
to sinusoidal functions multiplied by constants, 
as functions of $\sqrt{2mE/N}$. 
   To get the data for this figure we used the quasi-one-dimensional 
system of $50$ hard-disks with (H,P) boundary conditions, 
and calculated the amplitudes 
$|\alpha'_{1}|$ and $|\tilde{\alpha}'_{1}|$ 
for different total energies $E$. 
   In Fig. \ref{figB5oscillampli} we fitted 
the amplitude $|\tilde{\alpha}'_{1}|$ for 
the mode $\langle \delta x_{j}^{(2N-3)} \rangle_{t}$ 
(circles) 
to a constant function $y=\xi$ (the broken line)
with a fitting parameter value $\xi\approx  0.179$, 
and the solid line is given by $y=\xi x$ using this value 
of $\xi$. 
   The amplitudes $|\alpha'_{1}|$ for the modes 
$\langle \delta x_{j}^{(2N-3)}/p_{xj} \rangle_{t}$ 
(triangles), 
and $\langle \delta y_{j}^{(2N-3)}/p_{yj} \rangle_{t}$ 
(squares) are reasonably on the line  $y=\xi x$, 
and these results support the relation 
(\ref{normaLyavec1}), and also suggest that the amplitude  
$|\tilde{\alpha}'_{k}|$ for the mode 
$\langle \delta x_{j}^{(2N-3)} \rangle_{t}$ is 
independent of $\sqrt{2mE/N}$.   
   The amplitude $|\alpha'_{k}|$ for 
$\langle \delta x_{j}^{(2N-3)}/p_{xj} \rangle_{t}$ (triangles) 
and $\langle \delta y_{j}^{(2N-3)}/p_{yj} \rangle_{t}$ (squares)
in Fig. \ref{figB5oscillampli} almost coincide with each other, 
and it gives support to the claim that the coefficients 
$\alpha'_{k}$ (in the first term) on the right-hand side 
of Eqs. (\ref{LyapuOscilWPx1}) and (\ref{LyapuOscilWPy1}) 
coincide.

   It may be noted that the normalization procedure 
in the Benettin's algorithm, which we used to calculate 
the Lyapunov exponents and vectors in this paper, 
requires the normalization of Lyapunov vectors including 
both their spatial part and momentum part, but 
in the above argument we assumed the normalizability 
of the spatial part only of the Lyapunov vectors. 
   This can be justified by the fact that 
as shown in Sec. \ref{LyapunovModesMomentumComponents} 
the spatial part and 
momentum part of Lyapunov vectors constructing 
Lyapunov modes show almost the same mode structure, 
so each of them should be independently normalizable.

\subsection{Spatial node structures of the Lyapunov modes and reflections in the hard-walls}
\label{SpatialNodeStructure}

   The spatial node structure of the Lyapunov modes can be 
explained using the collision rule for particles 
with hard-walls. 

   For (H,P) boundary condition  
the particle collisions with the hard-walls 
in the $x$-direction cause a change in 
the sign of the $x$-component of the momentum 
with the remaining components of the phase space 
vector unchanged: 
\begin{eqnarray}
x_{j} &\rightarrow&   x_{j}, \\ 
y_{j} &\rightarrow&   y_{j}, \\ 
p_{xj} &\rightarrow& - p_{xj}, \\ 
p_{yj} &\rightarrow&   p_{yj}. 
\end{eqnarray}
%
   Similarly, in this type of collision the $x$-components 
of the Lyapunov vector change their signs 
while the remaining components unchanged:
\begin{eqnarray}
\delta x_{j} &\rightarrow& - \delta x_{j}, \\
\delta y_{j} &\rightarrow&   \delta y_{j}, \\
\delta p_{xj} &\rightarrow& - \delta p_{xj}, \\
\delta p_{yj} &\rightarrow&   \delta p_{yj}. 
\end{eqnarray}
%
Note that in the $x$ components of Lyapunov vector 
$\delta x_{j}$ changes its sign as well as 
$\delta p_{xj}$, which is different from the phase space vector.

   The important point is that a system with hard-wall boundaries
is equivalent to an infinite system generated by reflecting 
the positions and velocities of all particles (in the hard wall)
and by changing the signs of all 
$x$-components of the Lyapunov vectors 
at the hard-wall. That is explicitly incorporating  the
reflection symmetries 
for the phase space vector and the Lyapunov vector 
at hard walls.  
   If the modes of the entire system are connected smoothly 
sinusoidal functions at the hard walls, then this condition requires that 
the mode for the quantity $\delta x_{j}$ has 
a node at a hard wall, because it changes sign there. 
   On the other hand, the quantities 
$\delta x_{j}/p_{xj}$  and $\delta y_{j}/p_{yj}$ 
do not change their signs at hard walls, so these modes 
should have anti-nodes at hard walls. 
   These results explain the spatial node structures 
shown in Fig. \ref{figB4modeoscil2dconto}. 
   The spatial node structure of the stationary Lyapunov modes 
in  $\delta y_{j}$ corresponding to the one-point steps 
can be explained in this way. Because $\delta y_{j}$ varies 
sinusoidally and must satisfy the reflection symmetry, it must
be either a node (if the sign changes) or an anti-node 
(if the sign is invariant). Hence, in this case the 
Lyapunov mode in $\delta y_{j}$ should have an anti-node 
at the hard-walls.

\subsection{Lyapunov modes in momentum components of Lyapunov vectors}
\label{LyapunovModesMomentumComponents}

   So far, we have discussed only the spatial components 
of the Lyapunov vectors. 
   In this subsection we discuss briefly the 
Lyapunov modes appearing in the momentum parts of
Lyapunov vectors. 

   One of the few differences between spatial 
and momentum components of Lyapunov vectors is that 
the amplitudes of the momentum components 
are often much smaller than those of the corresponding 
spatial components \cite{Tan03b}. 
   This makes Lyapunov modes for the momentum parts of the 
Lyapunov vectors less clear than the 
corresponding spatial components. 
   However, basically the structure of the Lyapunov mode 
for the momentum part of the Lyapunov vector is quite similar 
to the corresponding spatial component. 
   For this reason, in this subsection 
we omit a detailed discussion of the 
phase relations of multiple Lyapunov modes for the 
momentum parts of Lyapunov vectors, and just show that 
there are certain modes structures in the momentum components 
of Lyapunov vectors corresponding to the Lyapunov steps.

\begin{figure}[!htb]
\vspfigA
\caption{ 
      Global time averages $\langle \delta p_{yj}^{(n)} \rangle$ 
   for the Lyapunov exponents corresponding to the 
   first, second and third one-point step of the Lyapunov spectrum 
   [$n = 198$ (circles),$195$ (triangles) and $192$ (squares)]
   as functions of the normalized global time average 
   $\langle x_{j} \rangle /L_{x}$ of the position of 
   the $j$-th. 
   }
\vspfigB
\label{figB6ModAvGloDpy}\end{figure}  
   
   Figure \ref{figB6ModAvGloDpy} shows mode structure  
of $\delta p_{yj}^{(n)}$ corresponding to 
the first three one-point steps 
($n = 198$,$195$ and $192$) as functions of the normalized 
particle positions. 
   This structure is stationary in time, so we took their 
global time average over $200N$ collisions, using
the notation $\langle\cdots\rangle$ for this global 
time-average without the suffix $t$. 
   These mode structures are stationary in time, and 
are similar to the ones for 
the corresponding spatial part $\delta y_{j}^{(n)}$  
discussed in Sect. \ref{StationaryLyapunovModes}.

\begin{figure}[!htb]
\vspfigA
\caption{ 
      Contour plots of the local time averages of 
   $\langle \delta p_{xj} \rangle_{t}$,   
   $\langle \delta p_{xj}/p_{xj} \rangle_{t}$, 
   and  
   $\langle \delta p_{yj}/p_{yj} \rangle_{t}$  
  for the first exponent (197) of the first two-point step 
  as a function of the collision number $n_{t}$ and 
   the normalized position of the $j$-th 
   particle (in the same collision number interval 
   $[535200,569600]$). 
      The dotted lines, solid lines, and broken lines 
   are contour lines at the levels $-0.0018$, $0$, and $+0.0018$, 
   respectively. 
   }
\vspfigB
\label{figB7ModMomen04conto}\end{figure}  

   Figure \ref{figB7ModMomen04conto} shows contour plots 
of time-oscillating 
Lyapunov modes for 
$\langle \delta p_{xj}^{(197)} \rangle_{t}$,  
$\langle \delta p_{xj}^{(197)}/p_{xj} \rangle_{t}$, 
and  
$\langle \delta p_{yj}^{(197)}/p_{yj} \rangle_{t}$ 
as functions of the collision number $n_{t}$ and 
the normalized local time average 
$\langle x_{j} \rangle_{t}/L_{x}$
in the first two-point step. 
   We used the same collision number interval 
$[535200,569600]$ in Fig. \ref{figB7ModMomen04conto} 
as in Fig. \ref{figB4modeoscil2dconto}. 
   The mode structures in Figs. 
\ref{figB7ModMomen04conto} (a), (b) and (c) 
are almost the same 
as Figs. \ref{figB4modeoscil2dconto}(a), (c) and (e) 
for the corresponding spatial components 
$\langle \delta x_{j}^{(197)} \rangle_{t}$,  
$\langle \delta x_{j}^{(197)}/p_{xj} \rangle_{t}$, 
and  
$\langle \delta y_{j}^{(197)}/p_{yj} \rangle_{t}$, 
respectively, 
although their oscillating amplitudes are much smaller 
than those of the corresponding spatial components. 

   The spatial mode structures of 
the momentum components of Lyapunov vectors 
are explained by the same reflection property at hard-walls,  
which was discussed in the previous subsection 
\ref{SpatialNodeStructure}.

\subsection{Particle number dependence of the oscillating periods}
\label{NpDepenOscillLyaMod}

   In Sect. \ref{TimeOscillatingLyapunovModes} 
we showed that the quantities 
$\langle \delta x_{j}^{(n)} \rangle_{t}$, 
$\langle \delta x_{j}^{(n)}/p_{xj} \rangle_{t}$, and 
$\langle \delta y_{j}^{(n)}/p_{yj} \rangle_{t}$ corresponding 
to the two-point Lyapunov steps show 
time-oscillating behavior. 
   Now we consider how the time-oscillating period 
of those Lyapunov modes depends on the number of particles $N$ 
for the quasi-one-dimensional system. 

   We evaluate the collision number interval for the time-oscillation 
of Lyapunov modes as follows. 
   As shown in the proceeding subsection 
\ref{TimeOscillatingLyapunovModes}, the Lyapunov modes 
related to the quantity $\langle \delta x_{j}^{(n)} \rangle_{t}$ 
(the quantities $\langle \delta x_{j}^{(n)}/p_{xj} \rangle_{t}$ 
and $\langle \delta y_{j}^{(n)}/p_{yj} \rangle_{t}$) 
have an anti-node in the middle (at the end) of the system in the 
$x$-direction ($n=2N-3$ and $2N-4$) for (H,P) boundary condition. 
   Using this properties we took 6 data points for the 
quantity $\langle \delta x_{j}^{(n)} \rangle_{t}$ 
(the quantities $\langle \delta x_{j}^{(n)}/p_{xj} \rangle_{t}$ 
and $\langle \delta y_{j}^{(n)}/p_{yj} \rangle_{t}$) 
($n=2N-3$) 
in the middle (at the end) of the system with (H,P) 
boundary condition. 
   These data are fitted to a sinusoidal function 
$y = a \sin\{(2\pi x/ T_{lya}) + b\}$ with 
fitting parameters $a$, $b$, and $T_{lya}$, 
which leads to a numerical estimation of the period
$T_{lya}$ of the time-oscillation of the Lyapunov modes. 
   The collision number interval $T_{lya}$ can be translated 
into a real time interval by multiplying by the 
mean free time $\tau$, if necessary. 

\begin{figure}[!htb]
\vspfigA
\caption{
      Particle number ($N$) dependence of the period 
   $T_{lya}$ of the time-oscillations 
   of $\langle \delta x_{j}^{(2N-3)} \rangle_{t}$   (circles), 
   $\langle \delta x_{j}^{(2N-3)}/p_{xj} \rangle_{t}$ (triangles), 
   and $\langle \delta y_{j}^{(2N-3)}/p_{yj} \rangle_{t}$ (squares) 
   in the quasi-one-dimensional system with (H,P) boundary condition. 
      The data is fitted to the function $y=\alpha + \beta x^2$ 
   with the fitting parameters $\alpha$ and $\beta$. 
      The inset: Particle number dependence of the mean free time 
   $\tau$ with a fitting function $y=\gamma/x$ with the 
   fitting parameter $\gamma$. 
   }
\vspfigB
\label{figB8NdepenMode}\end{figure}  

   Figure \ref{figB8NdepenMode} is the graph of the period
$T_{lya}$ of the time-oscillations  
of $\langle \delta x_{j}^{(197)} \rangle_{t}$ (circles), 
$\langle \delta x_{j}^{(197)}/p_{xj} \rangle_{t}$ (triangles), 
and $\langle \delta y_{j}^{(197)}/p_{yj} \rangle_{t}$ (squares) 
in the quasi-one-dimensional system with (H,P)  boundary condition, 
as functions of the number of particles $N$. 
   Spatial and temporal behavior of these quantities has already been 
shown in Fig. \ref{figB3modeoscil3d}(a), (b), and (d) for $N=100$. 
   Fig.  \ref{figB8NdepenMode} shows that the 
three time-oscillations associated with  
$\langle \delta x_{j}^{(197)} \rangle_{t}$, 
$\langle \delta x_{j}^{(197)}/p_{xj} \rangle_{t}$, 
and $\langle \delta y_{j}^{(197)}/p_{yj} \rangle_{t}$ all have 
the same period.  
   In Fig. \ref{figB8NdepenMode} the data is fitted to a 
quadratic function $y = \alpha + \beta x^{2}$ with 
the fitting parameter values $\alpha \approx 17.9$ 
and $\beta \approx 1.65$
%
   The inset to this figure shows 
the mean free time $\tau$ as a function of the number of particles $N$. 
   The $N$ dependence of $\tau$ is nicely fitted to 
the function $y = \gamma / x$ with 
the value $\gamma \approx 1.91$ of the fitting parameter $\gamma$. 
   Noting that the period, in real time, is given by 
$\tau T_{lya}$  approximately, 
these results suggest that the period 
of the Lyapunov modes is proportional to the number of particles $N$, 
namely the system size. 

\begin{figure}[!htb]
\vspfigA
\caption{
      The quantity $L_{x}/(\tau \bar{T}_{acf})$ as a function 
   of the number of particles $N$. 
      Here, $\bar{T}_{lya}$ is the collision number interval 
   given by the average of the three collision number intervals 
   $T_{lya}$ for time-oscillations of the Lyapunov vector components 
   $\langle \delta x_{j}^{(2N-3)} \rangle_{t}$, 
   $\langle \delta x_{j}^{(2N-3)}/p_{xj} \rangle_{t}$, 
   and $\langle \delta y_{j}^{(2N-3)}/p_{yj} \rangle_{t}$ 
   in the first two point step of the Lyapunov spectra. 
      The line is given by $y=1$, which is the thermal velocity 
    $\sqrt{E/(MN)}$. 
   }
\vspfigB
\label{figB9NdepenModVel}\end{figure}  

   Now, we investigate the time-oscillating period 
of the Lyapunov modes in a different way.   
   Figure \ref{figB9NdepenModVel} shows the quantity 
$L_{x}/(\tau \bar{T}_{acf})$ using the system length $L_{x}$, 
the mean free time $\tau$ and 
the averaged time-oscillating periods $\tau \bar{T}_{acf}$ 
of the Lyapunov modes in 
$\langle \delta x_{j}^{(2N-3)} \rangle_{t}$, 
$\langle \delta x_{j}^{(2N-3)}/p_{xj} \rangle_{t}$, 
and $\langle \delta y_{j}^{(2N-3)}/p_{yj} \rangle_{t}$ 
in the first two-point 
step of the Lyapunov spectrum. 
   This figure suggests that this quantity is almost $1$ 
independent of the particle number $N$, therefore equal to 
the thermal velocity $\sqrt{E/(MN)}$. 

\section{Auto-Correlation Functions}
\label{AutoCorrelationFunctions}

   In this section we discuss another property 
of the quasi-one-dimensional system, namely the time-oscillation 
behavior of the momentum auto-correlation function. 
   This is a typical measure of the collective behavior 
of many-particle systems. 
   We connect this behavior with the time-oscillating behavior 
of the Lyapunov modes, suggesting that the time-oscillation 
of the Lyapunov modes can also be regarded as a collective mode. 

   We calculate numerically the auto-correlation 
functions $C_{\eta} (t)$ for the $\eta$-component of the momentum 
based on the normalized expression 
$C_{\eta}(t) \equiv \tilde{C}_{\eta}(t)/\tilde{C}_{\eta}(0)$, 
in which $\tilde{C}_{\eta}(t)$ is defined by 

\begin{eqnarray}
   \tilde{C}_{\eta}(t) &\equiv& 
      \lim_{T\rightarrow +\infty} 
      \frac{1}{(N_{2}-N_{1}+1)T} 
   \nonumber \\
   && \times \sum_{j=N_{1}}^{N_{2}} 
   \int_{0}^{T} ds \; 
      p_{\eta j}(s+t) p_{\eta j}(s) , 
\label{AutoCorreFunct}\end{eqnarray}

\noindent $\eta =x$ or $y$.  
   Eq. (\ref{AutoCorreFunct}) includes a time-average and 
an average over some of particles (from the $N_{1}$-th particle 
to the $N_{2}$-th particle) in the middle of the system. 
   (Note that number the particles $1,2,\cdots,N$ from left to right  
in the system, as shown in Fig. \ref{figA1qua1dim}.)
   In actual calculations we choose $N_{1}=[(N+1)/2]-5$ 
and $N_{2}=[(N+1)/2]+5$ 
with $[x]$ as the integer part of the real number $x$. 
   This means that we take into account only 11 particles 
in the middle of the system in the calculation 
of the auto-correlation function $C_{\eta}(t)$.  
(In this paper we consider the case $N \geq 40 > 11$.)
   It should be noted that using (H,P)  boundary condition 
the auto-correlation function for particles near hard-walls 
are different from the ones for particles in the middle of the system, 
as discussed in Appendix \ref{TCFEachParticle}. 
   Especially, the momentum auto-correlation function 
of particles near hard-walls do not show 
clear time-oscillating behavior. 
   To get the clearest time-oscillating behavior for the 
auto-correlation function $C_{\eta}(t)$ and to get less 
hard-wall boundary condition effects, we exclude the 
auto-correlation functions of particles near hard-walls 
in the calculation of $C_{\eta}(t)$.
   
   If the system is ergodic, the value of the auto-correlation 
function  (\ref{AutoCorreFunct}) will be independent of the initial 
condition.    
   To get the results for the auto-correlation function 
in this section we take a time-average 
of the auto-correlation function over more than $2 \times 10^{6}$ 
collisions. 
   For convenience, in figures the auto-correlation functions  
are shown as functions of the correlation number $n_{t}$, but 
it can always be translated into the real time $t$ by multiplying by 
the mean free time $\tau$.

\subsection{Momentum auto-correlation functions and their direction dependence}
\label{MomentumAutoCorrelationFunctions}

   Figure \ref{figC1tcfCxCy} contains the momentum auto-correlation 
functions $C_{x}$ and $C_{y}$ for the momentum components 
in the $x$- and $y$-directions, respectively, 
as a function of the collision number $n_{t}$ 
in the quasi-one-dimensional system with $N=100$ 
and (H,P) boundary condition. 
   The main figure in Fig. \ref{figC1tcfCxCy} is a linear-linear plot 
of the auto-correlation functions $C_{x}$ and $C_{y}$, 
while its inset is a log-log plot of the graph of the absolute 
values $|C_{x}|$ and $|C_{y}|$ of the momentum auto-correlation 
functions. 
   In this system the mean free time is given 
by $\tau \approx 0.0188$. 
   From Fig. \ref{figC1tcfCxCy} it is clear that the momentum 
auto-correlation function has a strong direction dependence 
and shows a time-oscillating behavior in $C_{x}$.

\begin{figure}[!htb]
\vspfigA
\caption{ 
      The auto-correlation function $C_{x}$ and $C_{y}$ 
   for the $x$- and $y$-components of momentum, respectively, 
   as functions of the collision number $n_{t}$. 
      Main figure: Linear-linear plots of $C_{x}$ and $C_{y}$ 
   as functions of $n_{t}$. 
      The inset: Log-log plots of the absolute values 
   $|C_{x}|$ and $|C_{y}|$ of the auto-correlation functions 
   as functions of $n_{t}$. 
      Here, the broken line is a fit 
   of the graph of $|C_{x}|$ to an exponential function, 
   and the line is a fit of the graph of 
   $|C_{y}|$ to a $\kappa$-exponential function (defined by 
   Eq. (\ref{kexponential})). 
   }
\vspfigB
\label{figC1tcfCxCy}\end{figure}  

   In the beginning, the auto-correlation function $C_{x}$ for the 
$x$-component of momentum decays exponentially in time.  
   To show this point, in the inset to Fig. \ref{figC1tcfCxCy} 
we fitted the beginning of the graph of $|C_{x}|$ 
to an exponential function 

\begin{eqnarray}
   G_{1}(x) =\exp\{-\alpha' x\}
\label{ExponentFitting}\end{eqnarray}

\noindent with the fitting parameter value $\alpha' \approx 0.0385$. 

\begin{figure}[!htb]
\vspfigA
\caption{ 
      The time-oscillating part of auto-correlation function
   $C_{x}$  for the $x$-component of momentum 
   as functions of the collision number $n_{t}$. 
      Here, the line is the fit to the product of a sinusoidal and
  an  exponential function. 
   }
\vspfigB
\label{figC2TcfOscill}\end{figure}  

   The significant point about the auto-correlation function $C_{x}$ 
is its time-oscillating behavior. 
   To show this behavior explicitly we show Fig. \ref{figC2TcfOscill} 
as an enlarged graph of the time-oscillating part of $C_{x}$, 
which is already shown in Fig. \ref{figC1tcfCxCy}. 
   This time-oscillation accompanies a time decay, so we 
fitted this graph to the product of a sinusoidal and an 
exponential function $G_{2}(x)$, namely 

\begin{eqnarray}
   G_{2}(x) \equiv \mathcal{A} \; e^{-\beta' x} \sin 
   \left( \frac{2\pi}{T_{acf}} x + \xi \right)
\label{SinusoidalPower}\end{eqnarray}

\noindent with fitting parameters $\mathcal{A}$, $\beta'$, 
$T_{acf}$, and $\xi$. 
   The time oscillating part of the auto correlation function 
$C_{x}$ is nicely fitted to this function with the parameter 
values 
$\mathcal{A}\approx 0.0209$, 
$\beta' \approx  5.17 \times 10^{-5}$, 
$T_{acf} \approx 8.29\times 10^{3}$, and 
$\xi \approx 1.62$.
   This also gives us a way of numerically evaluating the 
oscillation period $T_{acf}$ of the auto-correlation function $C_{x}$. 
   We note the quasi-one-dimensional system shows a much 
clearer time-oscillating behavior of the momentum auto-correlation 
function than a fully two- (or three-) dimensional system. 
   One may ask whether the damping behavior of the envelope  
of time-oscillation of $C_{x}$ is best fitted to a power-law function, 
like the slow damping of the long time behavior of $C_{y}$,  
rather than to an exponential function as assumed in Eq. 
(\ref{SinusoidalPower}). 
   (Actually the data in Fig. \ref{figC2TcfOscill} is not sufficient to 
decide between exponential and power decay.) 
   This point is discussed further in Appendix 
\ref{EnvelopOscillation}. 

   On the other hand, the auto-correlation function $C_{y}$ for the 
$y$-component of momentum shows a significantly different behavior 
from $C_{x}$.
   This comes from the specific shape of the system and the boundary 
condition. 
   As shown in Fig. \ref{figC1tcfCxCy}, the damping of the momentum 
auto-correlation function $C_{y}$ 
in the $y$-direction is much slower than in the $x$-direction. 
   This is simply explained by the fact that in the 
quasi-one-dimensional system particle collisions occur mostly 
as head-on collisions, so that the change in the $y$-component 
of momentum in a collision can be much smaller than 
the change of the $x$-component of momentum. 
   In the inset to Fig. \ref{figC1tcfCxCy} the graph 
of the auto-correlation function for the $y$-component of momentum 
is fitted to the "$\kappa$-exponential function" $F_{\kappa}(x)$, 
which is defined by 

\begin{eqnarray}
   F_{\kappa}(x) \equiv \left[
      \sqrt{1+(\alpha'' \kappa x)^{2}} - \alpha'' \kappa x 
   \right]^{1/\kappa} ,
\label{kexponential}\end{eqnarray} 

\noindent in $x \geq 0$  with fitting parameters $\alpha''$ and 
$\kappa$. 
   In the collision number region shown 
in Fig. \ref{figC1tcfCxCy} the auto-correlation function $C_{y}$ 
in the $y$-direction is positive, so this fitting can be 
for $C_{y}$ as well as for $|C_{y}|$. 
   From the definition, in the limit as $\kappa = 0$ 
the function $F_{\kappa}(x)$ becomes the exponential function: 
$\lim_{\kappa \rightarrow 0} F_{\kappa}(x) = 
\exp\{-\alpha'' x \}$, noting $F_{\kappa}(0)= 1$ and 
$\partial F_{\kappa}(x) / \partial x 
= -\alpha'' F_{\kappa}(x)/ 
\sqrt{1+(\alpha''\kappa x)^{2}}$, so 
this function is a one-parameter deformation of the 
exponential function \cite{Kan01}.  
   The important properties of this function are 
that it is approximated by an exponential function at 
small $x$ and is approximately  
a power function in a large $x$.   

\begin{eqnarray}
   F_{\kappa}(x) \sim 
   \left\{ \begin{array}{ll} 
      e^{-\alpha'' x} \;\; 
         &\mbox{in} \;\;\; x <\!< 1  \\
      (2\kappa\alpha'' x)^{-1/\kappa} \;\; 
         &\mbox{in} \;\;\; x >\!> 1  
         \;\; \mbox{and} \;\; \kappa \alpha'' x > 0
   \end{array} \right.    
   \nonumber \\
\label{quasiexpo1}\end{eqnarray}

\noindent The fitting of the numerical data for the auto-correlation 
function $C_{y}$ to the $\kappa$-exponential function with 
parameter values $\alpha'' \approx 0.00358$ and 
$\kappa \approx 1.44$ is very satisfactory, 
and this implies that this auto correlation 
function decays exponentially in the beginning, 
(like the auto-correlation function 
in the $x$-direction), and decays 
as a power function after that, at least 
in the time scale shown in Fig. \ref{figC1tcfCxCy}.  
   (This does not mean that the auto-correlation function $C_{y}$ 
decays as a $\kappa$-exponential function in any time-scale. 
   See Appendix \ref{EnvelopOscillation} about 
the auto-correlation function 
$C_{y}$ at much longer time-scales than shown 
in Fig. \ref{figC1tcfCxCy}.)


\subsection{Particle number dependence of the auto-correlation function 
and a relation with the time-oscillation of the Lyapunov modes}

   We have shown the two kinds of time-oscillation behaviors 
in the quasi-one-dimensional system: one for the Lyapunov mode and 
another for the momentum auto-correlation function. 
   Now we show numerical evidence to connect these two behaviors. 
   
\begin{figure}[!htb]
\vspfigA
\caption{
     The period  $\bar{T}_{lya}$ (in collision numbers)
   of the time-oscillation of the Lyapunov mode 
   as a function of the period of time-oscillation of the 
   longitudinal momentum auto-correlation function $T_{acf}$.    
      Data points are obtained from numerical calculation 
   of the quasi-one-dimensional system with (H,P) boundary 
   condition for different numbers of particles 
   $N=40,50,60,\cdots,100$. 
      The line is given by the function $\bar{T}_{lya}=2 T_{acf}$. 
   }
\vspfigB
\label{figC3periodTcfLyamod}\end{figure}  

   Fig. \ref{figC3periodTcfLyamod} is the graph of the time-oscillating 
period $\bar{T}_{lya}$ of the Lyapunov modes as a function of 
the time-oscillating period $T_{acf}$ of the momentum auto-correlation 
function $C_{x}$. 
   Here, the time-oscillating periods are calculated for  
$N=40,50,60,\cdots,100$, and the time-oscillating 
periods $\bar{T}_{lya}$ are calculated as the average 
of the collision number interval $T_{lya}$ for 
time-oscillations of the Lyapunov vector components 
$\langle \delta x_{j}^{(2N-3)} \rangle_{t}$, 
$\langle \delta x_{j}^{(2N-3)}/p_{xj} \rangle_{t}$, 
and $\langle \delta y_{j}^{(2N-3)}/p_{yj} \rangle_{t}$ 
in the first two-point step of the Lyapunov spectra. 
   (As shown in Sect. \ref{NpDepenOscillLyaMod}, these 
three oscillating periods $T_{lya}$ of the Lyapunov modes 
take almost the same values.)
   In Fig. \ref{figC3periodTcfLyamod} we also show the line 
given by the function $\bar{T}_{lya}=2 T_{acf}$. 
   The numerical data for the time-oscillations 
in Fig. \ref{figC3periodTcfLyamod} is nicely fitted to the function, 

\begin{eqnarray}
   T_{lya} \approx 2 T_{acf} . 
\label{RelationT}\end{eqnarray}

\noindent This is the main result of this paper.

\vspfigAa
\begin{table*}[htbp]
\caption{
      Time-oscillating periods and decay factors for the Lyapunov modes 
   and the momentum auto-correlation functions. 
      Here, $N$ is the number of particles, $\tau$ is the mean 
   free time, and $\bar{T}_{lya}$ is the collision number interval 
   given by taking an average of the three collision number intervals 
   $T_{lya}$ of time-oscillations of the Lyapunov vector components 
   $\langle \delta x_{j} \rangle_{t}$, 
   $\langle \delta x_{j}/p_{xj} \rangle_{t}$, 
   and $\langle \delta y_{j}/p_{yj} \rangle_{t}$ 
   corresponding to the first two point step of the 
   Lyapunov spectra. 
      The parameter $\alpha'$ is given by fitting the 
   beginning of the longitudinal momentum auto-correlation 
   function $C_{x}$ as a function of the collision 
   number $n_{t}$ to an exponential 
   function $y=\exp\{-\alpha' x\}$. 
      The parameters $\beta'$ and $T_{acf}$ are given by 
   fitting the time-oscillating part of the same function 
   $C_{x}$ to the 
   function $y = \mathcal{A} \exp\{-\beta' x\} \sin 
   [(2\pi/T_{acf}) x + \xi ]$. 
      The parameters $\alpha''$ and $\kappa$ are given by fitting 
   the transverse momentum auto-correlation function $C_{y}$  
   as a function of the collision number $n_{t}$ 
   to the $\kappa$-exponential function 
   $y = [ \sqrt{1+(\alpha'' \kappa x)^{2}} - \alpha'' \kappa x 
   ]^{1/\kappa}$ (Eq. (\ref{kexponential})). 
   }
\vspace{0.2cm}
\begin{center}
\begin{tabular}{ccc|ccc|cc} \hline\hline
   & \multicolumn{2}{c|}{Lyapunov mode} 
   & \multicolumn{3}{c|}{$Cx$}  
   & \multicolumn{2}{c}{$Cy$} \\ 
    \makebox[1.5cm]{N}                                  %
    & \makebox[\widthtableA]{$\tau$} 
    & \makebox[\widthtableA]{$\bar{T}_{lya}\;\tau$}
    & \makebox[\widthtableA]{$\tau/\alpha'$}            %
    & \makebox[\widthtableA]{$\tau/\beta'$} 
    & \makebox[\widthtableA]{$T_{acf}\;\tau$} 
    & \makebox[\widthtableA]{$\tau/\alpha''$}             %
    & \makebox[\widthtableA]{$1/\kappa$} 
    \\ \hline
   $40$ & 0.0489  & 123.5
   & 0.493 & 74.2 & 63.0
   & 5.33 & 0.732 \\
   $50$ & 0.0380  & 154.7
   & 0.483 & 102.5 & 76.0
   & 5.26 & 0.716 \\ 
   $60$ & 0.0326  & 197.3
   & 0.505 & 156.1 & 96.6
   & 5.37 & 0.675 \\ 
   $70$ & 0.0275  & 223.1
   & 0.493 & 196.1 & 107.9
   & 5.26 & 0.608 \\ 
   $80$ & 0.0238  & 256.7
   & 0.487 & 204.5 & 125.0
   & 4.74 & 0.577 \\ 
   $90$ & 0.0210   & 283.3
   & 0.487 & 252.0 & 139.9
   & 4.89 & 0.624 \\ 
   $100$ & 0.0188  & 305.7
   & 0.488 & 362.7 & 155.5
   & 5.27 & 0.693 \\ 
   \hline\hline
\end{tabular}
\end{center}
\label{sumar}
\end{table*}
\vspfigBa

   In Table \ref{sumar}, we summarize, not only the values 
of the time-oscillating periods $\bar{T}_{lya} \tau$ and 
$\bar{T}_{acf}\tau$ of the Lyapunov modes and 
the momentum auto-correlation function in real time, 
but also the data about the $N$-dependences of the damping properties
of the auto-correlation functions $C_{x}$ and $C_{y}$. 
   They include the mean free time $\tau$, 
the exponential damping times $\tau/\alpha'$ and $\tau/\beta'$ 
(for $C_{x}$) and $\tau/\alpha''$ (for $C_{y}$), and the power 
$1/\kappa$ of the damping of $C_{y}$ at long time. 
   Here, values of $\alpha'$, $\beta'$ and $T_{acf}$ 
($\alpha''$ and $\kappa$) are derived by fitting the 
auto-correlation function $C_{x}$ ($C_{y}$) to
Eqs. (\ref{ExponentFitting}) and (\ref{SinusoidalPower}) 
(Eq. (\ref{kexponential})). 
   From this table it is clear that the exponential damping times 
$\tau/\alpha'$ and $\tau/\alpha''$, and the power $1/\kappa$ 
are almost independent of the particle number $N$. 
   On the other hand, the exponential damping time 
$\tau/\beta'$ of the time oscillation of the auto-correlation 
$C_{x}$ increases as $N$ increases. 
   (We have already discussed the $N$-dependence of the 
mean free time $\tau$ in Sec. \ref{NpDepenOscillLyaMod}.)

\subsection{Boundary condition effects}
\label{BoundaryConditionEffects}

   So far, we have concentrated into the quasi-one-dimensional system 
with hard-wall boundary conditions in the $x$-direction 
and periodic boundary conditions in the $y$-direction, 
namely the (H,P) boundary condition, for technical 
convenience in the analysis of the Lyapunov modes. 
   On the other hand, in Ref. \cite{Tan03a} we have already 
discussed and compared Lyapunov steps and modes 
in the different boundary conditions: 
the purely periodic boundary conditions (P,P), 
the purely hard-wall boundary conditions (H,H), 
and periodic boundary conditions in the $x$-direction 
and hard-wall boundary conditions in the $y$-direction (P,H) 
as well as the boundary condition (H,P). 
   In this section we carry out a similar discussion 
for the momentum auto-correlation functions $C_{x}$
in these different boundary conditions.  
   Figure. \ref{figC4boundaries} contains schematic illustrations 
of these boundary conditions. 

\begin{figure}[!htb]
\vspfigA
\caption{
      Schematic illustrations of the four boundary conditions 
   (P,P), (P,H), (H,P) and (H,H) used in quasi-one-dimensional systems. 
      Here, (P,P) is the purely periodic boundary conditions, 
   (P,H) is periodic boundary conditions in the $x$-direction and
   hard-wall boundary conditions in the $y$-direction, 
   (H,P) is hard-wall boundary conditions in the $x$-direction and 
   periodic boundary conditions in the $y$-direction, and  
   (H,H) is the purely hard-wall boundary conditions. 
      The dashed lines and the solid lines on the boundaries 
   represent periodic boundary conditions and hard-wall boundary 
   conditions, respectively. 
   }
\vspfigB
\label{figC4boundaries}\end{figure}  

   For meaningful comparisons between the different boundary conditions 
we use the same mass $m$ and radius $R$ for the particles, and the same 
number of particles ($N=50$). 
   Using the set of the lengths $(L_{x},L_{y})$ to define the size of the system 
in the $x$ and $y$ directions for (H,P) boundary conditions, then we use  
$(L_{x}-2R,L_{y})$ for (P,P) boundary conditions 
$(L_{x}-2R,L_{y}+2R)$ for (P,H) boundary conditions, and
$(L_{x},L_{y}+2R)$ for (H,H) boundary conditions. 
   This gives the same effective area for particles to move in each of the four systems. 
   This also means that the mean free time $\tau$ in 
these four types of the boundary conditions will be the same 
(Concrete numerical values of $\tau$ are given in Table 
\ref{sumarPeriodboundary}.).

\begin{figure}[!htb]
\vspfigA
\caption{ 
      Auto-correlation functions $C_{x}$ for the $x$-component 
   of the momenta as functions of the collision number $n_{t}$ 
   for the boundary conditions (P,P), (P,H), (H,P) and (H,H). 
      The systems are quasi-one-dimensional systems 
   consisting of 50 hard-disks.  We observe:     
      (a) Exponential decay region in the beginning of the
   damping of the auto-correlation function $C_{x}$ as a 
   linear-log plot. 
      The dotted line and the broken line are the fits 
   for the cases (P,P) and (H,P) and the cases (P,H) 
   and (H,H) to exponential functions, respectively. 
      (b) Time-oscillating region of the auto-correlation functions 
   $C_{x}$ as a linear-linear plot. 
      The four graphs of the auto-correlation functions $C_{x}$ 
   are fitted to the functions $ y = \mathcal{A} \; \exp\{-\beta' x\} 
   \sin \{ (2\pi/T_{acf}) x + \xi \}$ with 
   the fitting parameters $\mathcal{A}$, $\beta'$, 
   $T_{acf}$ and $\xi$. 
   }
\vspfigB
\label{figC5tcfBoundary}\end{figure}  

   Fig. \ref{figC5tcfBoundary} shows the auto-correlation 
functions $C_{x}$ for the $x$-component of the momenta in 
quasi-one-dimensional systems consisting of $50$ hard-disks 
with boundary conditions (P,P), (P,H), (H,P) and (H,H) as 
functions of the collision number $n_{t}$. 
   Here, Fig. \ref{figC5tcfBoundary}(a) is the beginning part of the 
autocorrelation functions $C_{x}$, and is given 
as a linear-log plot to show 
their exponential decay as straight lines. 
   In this figure the fits to the exponential function 
(\ref{ExponentFitting}) with the fitting parameter 
$\alpha'$ are given for the cases (P,P) and (H,P) 
and the cases (P,H) and (H,H) separately. 
   The dotted line is the fit for the cases  
(P,P) and (H,P) with the fitting parameter values 
$\alpha' \approx  0.0765$, and the broken line is 
for the cases 
(P,H) and (H,H) with the fitting parameter values 
$\alpha' \approx 0.0597$. 
   Fig. \ref{figC5tcfBoundary}(b) is the 
time-oscillating part of $C_{x}$ in the four different 
boundary conditions. 
   In this figure, each auto-correlation function is fitted 
to a function 
(\ref{SinusoidalPower}) with the fitting parameters $\mathcal{A}$, 
$\beta'$, $T_{acf}$, and $\xi$. 
   The values of these fitting parameters are 
$(\mathcal{A}, \beta', T_{acf}, \xi) 
\approx (0.0422, 0.00108, 1.00\times 10^{3}, 1.37)$
for (P,P), 
$(\mathcal{A}, \beta', T_{acf}, \xi) 
\approx (0.0447, 0.000803, 1.24\times 10^{3}, 1.33)$ 
for (P,H), 
$(\mathcal{A}, \beta', T_{acf}, \xi) 
\approx (0.0398, 0.000369, 2.04\times 10^{3}, 1.56)$ 
for (H,P), and 
$(\mathcal{A}, \beta', T_{acf}, \xi) 
\approx (0.0403, 0.000246, 2.53\times 10^{3}, 1.56)$ 
for  (H,H). 

   Figure \ref{figC5tcfBoundary}(a) shows that the boundary condition 
in the $y$-direction has a strong effect on the auto-correlation 
function $C_{x}$ even at short time.
   In the cases (P,P) and (H,P) the auto-correlation function 
$C_{x}$ shows faster exponential decay than for 
(P,H) and (H,H), as shown in the difference of the value of 
the fitting parameter $\alpha'$ for the exponential fitting 
function (\ref{ExponentFitting}). 
   In the collision number region shown in Fig. 
\ref{figC5tcfBoundary}(a), the effects of the boundary conditions 
in the $x$-direction in the auto-correlation function $C_{x}$ 
appear after showing their initial exponential decays, and  
the autocorrelation function $C_{x}$ for 
(P,P) (and (P,H)) decays faster than $C_{x}$ 
for (H,P) (and (H,H)). 
   
   On the other hand, Fig. \ref{figC5tcfBoundary}(b) shows 
that in all the boundary conditions the auto-correlation 
functions $C_{x}$ show time-oscillations, but with different 
oscillating periods $T_{acf}$. 
   In this figure we can recognize that the second peak of $C_{x}$ 
for (P,P) coincides with the first peak 
of $C_{x}$ for (H,P). 
   A similar coincidence appears in the second peak of $C_{x}$ 
for (P,H) and the first peak of $C_{x}$ in (H,H). 
   Actually the fitting parameter value of $T_{acf}$ for (P,P) ((P,H)) 
is approximately the half the value of $T_{acf}$ for (H,P) ((H,H)). 
   This can be simply explained by the fact that replacing the 
periodic boundary conditions with the hard-wall boundary 
conditions, it needs twice the time for a particle perturbation 
to come back to the same position.

\vspfigAa
\begin{table}[htbp]
\caption{
      The mean free time $\tau$, the averages of the longest 
   time-oscillating time period 
   $\bar{T}_{lya}\tau$ for the Lyapunov vector, and the 
   time-oscillating time period $T_{acf} \tau$ of the 
   longitudinal momentum auto-correlation 
   function $C_{x}$ for the different boundary 
   conditions (P,P), (P,H), (H,P) and (H,H) in a quasi-one-dimensional 
   system consisting of $50$ hard-disks. 
      $\bar{T}_{lya}$ is the average over three collision number intervals of $T_{lya}$. 
   Here, $T_{lya}$ is the average of the time-oscillations of the three components 
   $\langle \delta x_{j}^{(k)} \rangle_{t}$, 
   $\langle \delta x_{j}^{(k)}/p_{xj} \rangle_{t}$, 
   and $\langle \delta y_{j}^{(k)}/p_{yj} \rangle_{t}$ 
   for the first Lyapunov step which has 
   time-oscillating Lyapunov modes 
   ($k=2N-5$ for (P,P), $k=2N-2$ for (P,H), $k=2N-3$ for (H,P), 
   and $k=2N-1$ for (H,H)). 
   }
\vspace{0.2cm}
\begin{center}
\begin{tabular}{c|ccc} \hline\hline
    \makebox[\widthtableB]{Boundary}                                  
    & \makebox[\widthtableB]{$\tau$} 
    & \makebox[\widthtableB]{$\bar{T}_{lya}\;\tau$}
    & \makebox[\widthtableB]{$T_{acf}\;\tau$} 
    \\ \hline
   (P,P) & 0.0369 & 37.4 & 77.0 \\
   (P,H) & 0.0371 & 45.8 & 91.4 \\
   (H,P) & 0.0380 & 77.3 & 154.5 \\
   (H,H) & 0.0383 & 96.8 & 194.5 \\
%
   \hline\hline
\end{tabular}
\end{center}
\label{sumarPeriodboundary}
\end{table}
\vspfigBa

   Finally we show a relation between the time-oscillating periods 
of the Lyapunov mode and the momentum auto-correlation function 
for different boundary conditions. 
   In Table \ref{sumarPeriodboundary} we summarize the mean 
free time $\tau$, the time-oscillating time period 
$\bar{T}_{lya}\tau$ of the largest Lyapunov mode, and 
the the time-oscillating time period 
$T_{acf}\tau$ of the momentum auto-correlation function $C_{x}$ 
for the four kinds of the boundary conditions (P,P), (P,H), 
(H,P) and (H,H). 
   Here, the period $\bar{T}_{lya}$ is evaluated as the 
arithmetic average of the collision number intervals 
$T_{lya}$ for the quantities 
$\langle \delta x_{j}^{(k)} \rangle_{t}$, 
$\langle \delta x_{j}^{(k)}/p_{xj} \rangle_{t}$, 
and $\langle \delta y_{j}^{(k)}/p_{yj} \rangle_{t}$. 
   The Lyapunov indices $k$ are chosen from the Lyapunov 
exponents in the first Lyapunov step which has 
time-oscillating behavior of its Lyapunov modes. 
   Our result supports the conjecture that the relation (\ref{RelationT}) 
is satisfied for all boundary conditions (P,P), (P,H), 
(H,P) and (H,H).

\subsection{An explanation for the relation of time-oscillation periods of the Lyapunov mode and the momentum auto-correlation function}

   As we have shown, the relation $T_{lya} = 2T_{acf}$ 
(Eq. (\ref{RelationT})) between the largest time-oscillating period 
$T_{lya}$ of the Lyapunov modes and the time-oscillating period 
$T_{acf}$ of the momentum auto-correlation function is independent 
of the number of particles $N$ and the boundary conditions. 
   In this subsection we discuss a possible explanation 
for this relation, which is a physical argument   
rather than a strict mathematical proof. 

   We consider a momentum component $\tilde{p}_{x}(t)$, 
like the $x$-component of the momentum in the 
quasi-one-dimensional system, 
which shows a time-oscillating behavior in its 
auto-correlation function with a frequency $\omega_{acf}$:

\begin{eqnarray}
   \overline{\tilde{p}_{x}(t)^{*} \; \tilde{p}_{x}(0)} 
   \sim \phi(t) \; e^{i\omega_{acf} t}
\label{acfmomenexpl}\end{eqnarray}

\noindent where we use the notation $\overline{X(t)^{*}X(0)}$ 
for the auto-correlation function for any complex quantity $X(t)$
with complex conjugate $X(t)^{*}$. 
   Here $\phi(t)$ is the damping envelope of the 
auto-correlation function 
$\overline{\tilde{p}_{x}(t)^{*} \; \tilde{p}_{x}(0)}$, 
which can be an exponential decay, 
$\phi(t)\sim \exp\{-\tilde{\alpha}t\}$ with a 
positive constant $\tilde{\alpha}$ 
(See Appendix \ref{EnvelopOscillation}.). 

   On the other hand, we represent the 
momentum proportional and time-oscillating term 
in the Lyapunov vector, like the first terms on 
the right-hand sides of Eqs. (\ref{LyapuOscilWPx1}) 
and (\ref{LyapuOscilWPx2}),  as 

\begin{eqnarray}
   \delta\tilde{q}_{x} \sim 
   \psi_{1}(t) \tilde{p}_{x}(t) e^{i\omega_{lya}t}
\label{lyaveccompexpl}\end{eqnarray}

\noindent with a frequency $\omega_{lya}$, 
where $\psi_{1}(t)$ is the decay envelope 
of the amplitude of $\delta\tilde{q}_{x}$, and 
may show an exponential divergence (or contraction) 
following the corresponding Lyapunov exponent. 
   Now, we assume that if the quantity 
$\delta\tilde{q}_{x}$ oscillates persistently in time, 
then its auto-correlation function 
$\overline{\delta\tilde{q}_{x}(t)^{*} \; \delta\tilde{q}_{x}(0)}$ 
should oscillate in time 
with the same frequency $\omega_{lya}$, namely 

\begin{eqnarray}
  \overline{\delta\tilde{q}_{x}(t)^{*} \;\delta\tilde{q}_{x}(0)} 
  \sim \psi_{2}(t) e^{i\omega_{lya}t} 
\label{acflyavecexpl}\end{eqnarray}

\noindent with a new envelope function $\psi_{2}(t)$. 

   It follows from Eqs. (\ref{acfmomenexpl}), 
(\ref{lyaveccompexpl}) and (\ref{acflyavecexpl}) that

\begin{eqnarray}
   \psi_{2}(t) e^{i\omega_{lya} t}  
      &\sim& \psi_{1}(t)^{*}\;\psi_{1}(0) \;
      \overline{\tilde{p}_{x}(t)^{*} \; \tilde{p}_{x}(0)} 
      \; e^{-i\omega_{lya} t}
      \nonumber \\
   &\sim& \psi_{1}(t)^{*}\;\psi_{1}(0)  \; \phi(t) \;  
      e^{i (\omega_{acf} - \omega_{lya})t} ,
\nonumber\end{eqnarray}

\noindent which immediately leads to

\begin{eqnarray}
   \psi_{2}(t)   &\sim& \psi_{1}(t)^{*}\;\psi_{1}(0) \; \phi(t) \\
   \omega_{lya} &\sim& \omega_{acf} /2 \label{omegarelatexpl}.
\end{eqnarray}

\noindent The time-oscillating periods $T_{acf}$ and 
$T_{lya}$ of the momentum auto-correlation function and the Lyapunov 
mode are given by $T_{acf} \sim 2\pi/(\tau\omega_{acf})$ and 
$T_{lya}\sim 2\pi/(\tau\omega_{lya})$. 
   Using this point and  Eq. (\ref{omegarelatexpl}) 
we obtain our relation (\ref{RelationT}). 
   Note that the above explanation for $T_{lya} = 2T_{acf}$ 
is independent of the number of particles $N$ and the boundary 
conditions. 

\begin{figure}[!htb]
\vspfigA
\caption{
      The normalized auto-correlation function $C_{lya,x}^{(2N-3)}$
   for the longitudinal Lyapunov 
   vector component $\delta x_{j}^{(2N-3)}$ as a function 
   of the collision number $n_{t}$. 
      The system is a quasi-one-dimensional system consisting 
   $50$ particles with boundary condition (H,P). 
      The numerical data is well fitted to a sinusoidal function 
   multiplied by an exponential function. 
   }
\vspfigB
\label{figC6TcfLyaVec}\end{figure}  

   In the above explanation, the assumption 
(\ref{acflyavecexpl}) is crucial, so it may be useful to demonstrate 
this behavior numerically. 
   Figure \ref{figC6TcfLyaVec} shows that the auto-correlation 
function $C_{lya,x}^{(2N-3)}$ 
for the longitudinal Lyapunov vector component 
$\delta x_{j}^{(2N-3)}$ normalized by its initial value 
(about 0.0203) in an quasi-one-dimensional system of 
$50$ particles 
with (H,P) boundary condition. 
   In the auto-correlation function $C_{lya,x}^{(2N-3)}$ 
its mean value 
is subtracted, and an average over the 
auto-correlation functions of 11 particles in the middle 
of the system is taken. 
   Here, the Lyapunov index $n=2N-3$ of the Lyapunov vector component 
$\delta x_{j}^{(n)}$ is chosen so that the corresponding 
Lyapunov step is the first two-point step associated 
with a time-oscillating Lyapunov mode.     
   In Fig. \ref{figC6TcfLyaVec} the numerical data is fitted to 
a sinusoidal function multiplied by an exponential function, 
namely the function (\ref{SinusoidalPower}), with 
the fitting parameter values $\mathcal{A} \approx 0.967$, 
$\beta' \approx 1.03\times 10^{-5}$, 
$T_{acf} = \tilde{T}_{acf} \approx 4.12 \times 10^{3}$, 
and $\xi \approx 1.54$.  
   This time-oscillating period $\tilde{T}_{acf}$ 
for the auto-correlation function for the longitudinal 
Lyapunov vector component coincides almost exactly with the 
time-oscillating period $T_{acf} \approx 4.07\times 10^{3}$ 
of the corresponding Lyapunov mode. 
   This coincidence of the time-oscillating periods supports  
our assumption (\ref{acflyavecexpl}) \cite{noteC1}. 

\section{Conclusion and Remarks}
\label{ConclusionRemarks}

   In this paper we have discussed the relation between the 
wave-like structure of Lyapunov vectors 
and the time-oscillating behavior
of the momentum auto-correlation functions in  
quasi-one-dimensional many-hard-disk systems. 
   The quasi-one-dimensional system is a 
narrow rectangular system in which the $x$ components
of the particle positions remained in the same order.  
   This system was proposed as a many-particle system 
which shows clear stepwise structure of the Lyapunov spectrum 
(the Lyapunov steps) and wave-like structure of the 
associated Lyapunov vectors (the Lyapunov modes).  
   Using this system, we showed that there are two 
types of Lyapunov modes in the spatial 
and momentum components of the 
Lyapunov vectors corresponding to the two kinds of steps 
in the Lyapunov spectrum: one is stationary in time 
and the other involves a time-oscillation. 
   Here, the time-oscillating Lyapunov vectors 
consist of a simple time-oscillating part plus 
a momentum proportional time-oscillating part 
in the longitudinal components, 
while the transverse time-oscillating Lyapunov vectors 
consist of a momentum proportional time-oscillating part only. 
   We revealed the phase relation for these time-oscillating 
Lyapunov modes.
   It was shown that the system length divided by the 
time-oscillating period of the Lyapunov modes 
is independent of the number of particles at the same density, 
and is of order of the thermal velocity.  
   After discussing these wave-like structures 
of the Lyapunov vectors, 
we connected them to the time-oscillation of the momentum 
auto-correlation. 
   The time-oscillation of the auto-correlation function 
appears in the longitudinal component of the momentum,  
and its envelope decays exponentially in time. 
   The main point is that the largest time-oscillating period 
of the first time-oscillating Lyapunov modes 
is twice as long as the time-oscillating 
period of the momentum auto-correlation function. 
   We showed that this relation is independent 
of the number of particles and the boundary conditions 
(constructed from combinations of periodic and hard-wall 
boundary conditions). 
   A simple explanation is given for this relation. 
   It was also shown that the auto-correlation function for the 
transverse component of the momentum is nicely fitted to the 
$\kappa$-exponential function, implying that it decays 
exponentially at the beginning and decays as a power after that. 

   In this paper we considered mainly a specific boundary 
condition for the quasi-one-dimensional system: (H,P) 
hard-wall boundary conditions in the longitudinal direction 
and periodic boundary conditions in the transverse 
direction. 
   The system with this boundary condition 
exhibit a much clear wavelike-like structure 
of Lyapunov modes than the purely periodic boundary 
conditions (P,P), which is a big advantage for quantitative 
discussions of the Lyapunov modes. 
   Using the (H,P) boundary condition, the spatial translational 
invariance in the longitudinal direction is violated, 
and it leads to a different Lyapunov step structure 
and auto-correlation functions, compared with the (P,P)  
boundary conditions. 
   For example, in (H,P) 
the step widths of the Lyapunov spectrum are half of  
the ones in (P,P), 
and individual particles can have different 
momentum auto-correlation functions due to the back scattering 
effect of the hard-wall (See Appendix 
\ref{BoundaryConditionEffects}) while 
the momentum auto-correlation function is 
particle-independent for the (P,P) boundary condition. 
   However, as discussed in Ref. \cite{Tan03a} for 
the Lyapunov modes and in Sec. \ref{BoundaryConditionEffects} 
for the auto-correlation functions, 
there is a simple relation connecting the results obtained from different 
boundary conditions, so we can predict some of the results of 
the other boundary conditions from the results for (H,P). 

   The mode structure of Lyapunov vectors discussed in this paper 
is related to the structure of the Lyapunov vectors 
associated with zero-Lyapunov exponents. 
   As explained in the introduction of this paper, 
there are sets of Lyapunov vector components which take a 
constant value independent of the particle index, and that 
these quantities corresponding to the stepwise structure 
of the Lyapunov spectrum have wavelike structures. 
   These are connected with the spatial and time translational 
invariances and the energy and momentum conservation laws. 
   However we need to be careful when making a connection  
between the conservation laws (or the translation invariances) 
and the Lyapunov modes. 
   For example, in a system with hard-wall boundary conditions 
the spatial translational invariance is violated, but even 
in such systems the mode structure in the Lyapunov vector component 
$\delta x_{j}^{(n)}$ (or $\delta y_{j}^{(n)}$), can be observed.
   However, a scenario which suggests that  translational invariance is
only evident when it is observed in the zero-Lyapunov exponent modes,  
will not predict these observed longitudinal modes.  

   It should be noted that a time-dependence of the Lyapunov modes 
may not always appear as a time-oscillating behavior. 
   Ref. \cite{For04} claim that the spatial wave of the 
Lyapunov vector "moves" at a specific speed in the 
square system consisting of many hard-disks. 
   It is interesting to know how these different behaviors, 
one oscillating in time and another moving with a speed, 
can appear. 

   In some papers, an understanding of the Lyapunov modes 
was attempted based on an analogy with the hydrodynamic modes \cite{For04}. 
   Actually, in both cases the conservation laws 
like the total momentum conservation and the energy conservation 
play an essential role, and the longitudinal mode shows 
a time-dependent behavior. 
   However it is important to know that the deterministic nature 
of orbits also plays one of the essential roles in the Lyapunov modes 
and leads to momentum-proportional time-oscillating components 
of Lyapunov vectors, although such a characteristic 
does not appear explicitly in the hydrodynamic mode. 
   In this sense, it is still an open question to see how hydrodynamic modes,
which have no concept of a phase space trajectory, can incorporate
time translational invariance.

   From results of this paper, it is suggested that 
there is a connection between existence of the 
stepwise structure of Lyapunov spectra and the 
time-oscillations of momentum auto-correlation functions. 
   It is well known that the stepwise structure of the 
Lyapunov spectra appears clearly in rectangular systems 
rather than in square systems at the same density. 
   It is possible to get a similar result for 
the time-oscillation of the momentum auto-correlation 
function? 
   For example, in a square system with a small number 
of hard-disks we cannot observe the stepwise structure 
of the Lyapunov spectrum, and in this case the time-oscillation 
of the momentum auto-correlation function does not appear. 
   Therefore, the time-oscillations of the 
auto-correlation function may be useful to 
understand the condition for existence of the Lyapunov steps 
and modes. 
   In this sense, for example, it may be interesting 
to investigate the time-correlation function in systems 
with soft-core particle interactions 
in which the observation of the Lyapunov steps is much harder, 
and less direct than in systems with hard-core interactions.

\section*{Acknowledgements}

   One of the authors (T.T.) thanks to A. Aliano for suggesting the 
$\kappa$-exponential function. 
   We are grateful for the financial support for this work 
from the Australian Research Council. 
   One of the authors (T. T.) also appreciates the financial 
support by the Japan Society for the Promotion Science.

\appendix
\section{Momentum Auto-Correlation Function of Individual Particle}
\label{TCFEachParticle}

   In this appendix we discuss the momentum auto-correlation 
function of individual particles in the quasi-one-dimensional system 
with hard-wall boundary conditions in the longitudinal direction and 
periodic boundary conditions in the transverse direction 
(the boundary condition (H,P)). 
   Different from purely periodic boundary 
conditions, the hard-wall boundary in  
(H,P) violates the translation invariance 
in the longitudinal  $x$-direction, 
and this implies that different particles can have different momentum 
auto-correlation functions. 
   
   We introduce the auto-correlation function  $c_{\eta}^{(j)}(t)$ 
of the $j$-th particle in the $\eta$-direction at time $t$ 
($j=1,2,\cdots,N$ and $\eta=x,y$) based on the normalized expression 
$c_{\eta}^{(j)}(t) \equiv \tilde{c}_{\eta}^{(j)}(t) 
/\tilde{c}_{\eta}^{(j)}(0)$ with $\tilde{c}_{\eta}^{(j)}(t)$ defined by 

\begin{eqnarray}
   \tilde{c}_{\eta}^{(j)}(t) \equiv 
      \lim_{T\rightarrow +\infty} 
      \frac{1}{T} \int_{0}^{T} ds \; 
      p_{kj}(s+t) p_{kj}(s) .
\label{AutoCorreFunctEachParti}\end{eqnarray}

\noindent Using this quantity $\tilde{c}_{\eta}^{(j)}(t)$, 
the auto-correlation function $\tilde{C}_{\eta}(t)$ 
defined by Eq. (\ref{AutoCorreFunct}) is simply given by 
$\tilde{C}_{\eta}(t) = [1/(N_{2}-N_{1}+1)]\sum_{j=N_{1}}^{N_{2}} 
\tilde{c}_{\eta}^{(j)}(t)$. 
   In this appendix we show graphs of $\tilde{c}_{\eta}^{(j)}$ 
as a function of the collision number $n_{t}\approx t/\tau$ 
in the quasi-one-dimensional system consisting of 
$100$ hard-disks. 
   We number the particles
$1,2,\cdots,N$ from the left to right, 
as shown 
in Fig. \ref{figA1qua1dim}, so, for example, the $1$st and $N$-th 
particles are closest to the hard-walls.

\begin{figure}[!htb]
\vspfigA
\caption{
      Auto-correlation functions for $c_{x}^{(1)}$, $c_{x}^{(20)}$ 
   and $c_{x}^{(50)}$ for the $x$-components of momenta of 
   the $1$st, $20$-th, and $50$-th particle, respectively, as 
   functions of the collision number $n_{t}$. 
      The system is a quasi-one-dimensional system 
   of $100$ hard-disks with (H,P) boundary condition.   
      The inset: Enlarged graphs 
   in the small magnitude part of the auto-correlation functions. 
   }
\vspfigB
\label{figF1TcfEachPartiLong}\end{figure}  

   The first important point of the individual 
auto-correlation functions is that 
the time-oscillation 
in the $x$-direction is weak for particles near the walls. 
   This is shown in Fig. \ref{figF1TcfEachPartiLong},  
in which we plot the auto-correlation function $c_{x}^{(j)}$ 
for the $j$-th particle in the $x$-direction for 
$j=1$ (the nearest particle to the left hard-wall), 
$j=50$ (the particle most distant from the hard-walls), 
and $j=20$ (a particle between these two extremes). 
   In Fig. \ref{figF1TcfEachPartiLong}, 
the main figure is the full data for these auto-correlation 
functions, and its inset is an enlarged graph 
to emphasize the time-oscillating part. 
   This figure shows that we cannot recognize a time-oscillating 
behavior in the auto-correlation function $c_{x}^{(1)}$ of 
the particle nearest to the hard-wall, although a clear 
time-oscillation can be recognized in $c_{x}^{(50)}$ 
of the particle in a middle of the system. 
   We can see a time-oscillating behavior in the 
auto-correlation function $c_{x}^{(20)}$, but its amplitude 
is smaller than that of $c_{x}^{(50)}$. 
   The positions of the nodes of the time-oscillations 
of $c_{x}^{(20)}$ and $c_{x}^{(50)}$ almost coincide with each other.

\begin{figure}[!htb]
\vspfigA
\caption{
      Exponential decay region of the auto-correlation functions 
   $c_{x}^{(1)}$ (circles), 
   $c_{x}^{(2)}$ (triangles), 
   and $c_{x}^{(50)}$ (squares) 
   for the $x$-components of momenta,  
   as functions of the collision number $n_{t}$ on a 
   linear-log plot. 
      The solid line (the dotted-broken lines) 
   is a fit of the graph of $c_{x}^{(1)}$ (the graphs 
   of $c_{x}^{(2)}$ and $c_{x}^{(50)}$) to an 
   exponential function.    
      Similar graphs are given for the auto-correlation functions 
   for the $y$-component of momenta: 
   $c_{y}^{(1)}$ (inverted triangles), 
   $c_{y}^{(2)}$ (pluses), 
   and $c_{y}^{(50)}$ (crosses). 
      The broken line (the dotted line) 
   is a fit of the graph of $c_{y}^{(1)}$ (the graphs 
   of $c_{y}^{(2)}$ and $c_{y}^{(50)}$) to an 
   exponential function.    
   }
\vspfigB
\label{figF2TcfEachPartiShort}\end{figure}  

   Another difference between individual particle 
auto-correlation functions 
appears in the short time scale. 
   Figure \ref{figF2TcfEachPartiShort} shows the momentum 
auto-correlation functions $c_{\eta}^{(j)}$, $j=1,2$ and $50$, 
as functions of the collision number $n_{t}$ 
showing the initial damping behavior ($\eta=x,y$). 
   These auto-correlation functions show an exponential 
decay, which we present as a linear-log plot 
(straight lines imply exponential decay).  
   This figure shows that the $x$-component ($y$-component) 
of the auto-correlation function of the particle nearest 
the hard-wall decays faster (slower) than 
the ones of other particles, while in any of them 
the damping behavior is nicely fitted to an exponential function. 
   (In Fig. \ref{figF2TcfEachPartiShort} the graphs are 
fitted to the exponential function (\ref{ExponentFitting}) 
with the fitting parameter values 
$\alpha' \approx 0.0457$ for $c_{x}^{(1)}$ (solid line), 
$\alpha' \approx 0.0369$ for $c_{x}^{(2)}$ and $c_{x}^{(50)}$ 
(dotted-broken line), 
$\alpha' \approx 0.00184$ for $c_{y}^{(1)}$ (broken line), and 
$\alpha' \approx 0.00368$ for $c_{y}^{(2)}$ and $c_{y}^{(50)}$ 
(dotted line) ). 
   This difference may come from the different types of collisions 
experienced.
   For the particle nearest the wall, half of the collisions will be with the wall 
and the other half with the neighboring particle. 
  The $x$-component of the momentum is drastically changed 
(namely it changes the sign of $p_{xj}$), so it may cause 
a faster decay of the autocorrelation function in the $x$-direction 
than for other particles. 
   On the other hand, collisions with the wall effect the $y$-component 
of the momentum much less, because it is invariant under wall collisions, 
and does not cause loss 
of memory of $p_{yj}$, and this can also explain a 
slower decay of the momentum auto-correlation for the $1$st 
(and $N$-th) particle in the $y$-direction compared to other particles.

\begin{figure}[!htb]
\vspfigA
\caption{
      The negative region of auto-correlation functions 
   $c_{x}^{(1)}$, $c_{x}^{(3)}$, $c_{x}^{(5)}$, 
   and $c_{x}^{(10)}$ for the $x$-components of momenta,  
   as functions of the collision number $n_{t}$ as a linear-linear plot. 
      The inset: Absolute values of the same auto-correlation 
   functions as functions of $n_{t}$ as a log-log plot. 
   }
\vspfigB
\label{figF3TcfEachPartiMiddleA}\end{figure}  

   Another point of difference of the auto-correlation functions 
of individual particles is a negative region which appears 
after their initial exponential decay. 
   It may be meaningful to mention that a negative region of 
momentum auto-correlation function has drawn attention in some previous works 
\cite{Rah64,Ber66,Zwa70}. 
   To discuss such a negative region, 
in Fig. \ref{figF3TcfEachPartiMiddleA} 
we show the collision number $n_{t}$ dependence of the 
auto-correlation functions 
$c_{x}^{(1)}$, $c_{x}^{(3)}$, $c_{x}^{(5)}$, 
and $c_{x}^{(10)}$. 
   To emphasize such a negative region of the auto-correlation 
function, we show a log-log plot of the absolute values of 
the same quantities as functions of $n_{t}$ in 
the inset to Fig. \ref{figF4TcfEachPartiMiddleB}. 
   As shown in this figure, a negative region of the auto-correlation 
functions appears after the initial exponential decay and before the  
time-oscillates appear. 
   The collision number (or time) at the bottom of this negative 
region of the auto-correlation function increases, and the amplitude 
of the bottom decreases, as the particle is further from the hard-wall 
(namely as the particle index $j$ in $c_{x}^{(j)}$ increases from $j=1$ to $10$ in 
Fig. \ref{figF4TcfEachPartiMiddleB}). 
   This phenomenon can be explained by the backscattering effect 
of the hard-walls. 
   Such a back scattering effect is stronger 
(so the amplitude of the negative region is stronger) 
in a particle closer to a hard-wall, 
   As well, the time interval to react to the presence of the wall 
is longer (so the time at the bottom of the negative region is later) 
in a particle far from the hard-wall. 
   This kind of behavior is not observed in a system in which the boundary conditions 
in the $x$-direction are periodic. 

\begin{figure}[!htb]
\vspfigA
\caption{
      The region before the start of the time-oscillation 
   of auto-correlation functions for 
   $c_{x}^{(10)}$, $c_{x}^{(20)}$, $c_{x}^{(35)}$  
   and $c_{x}^{(50)}$ 
   as functions of the collision number $n_{t}$. 
   }
\vspfigB
\label{figF4TcfEachPartiMiddleB}\end{figure}  

   After such a negative region of the auto-correlation function 
the time-oscillating part appears. 
   Figure \ref{figF4TcfEachPartiMiddleB} shows  
the collision number $n_{t}$ dependence of the 
auto-correlation functions for 
$c_{x}^{(10)}$, $c_{x}^{(20)}$, $c_{x}^{(35)}$ 
and $c_{x}^{(50)}$ 
in the collision number region before the time-oscillation 
of the auto-correlation starts (about $n_{t} \approx 6000$ 
in Fig. \ref{figF4TcfEachPartiMiddleB}). 
   The negative peak of the auto-correlation function 
(discussed in the previous paragraph and 
indicated by the arrows in Fig. \ref{figF4TcfEachPartiMiddleB}) 
moves to a longer collision number $n_{t}$ as the particle 
index $j$ increase from $j=10$ to $35$ in Fig.  
\ref{figF4TcfEachPartiMiddleB}. 
   On the other hand, the time oscillation 
of the auto-correlation function starts from about 
$n_{t} \approx 6000$ which is independent of the particle index, 
although the amplitude of the time-oscillation is large 
for a particle far from the hard-walls. 
   Moreover the time-oscillating period of the 
auto-correlation function is almost independent of the particle index. 
   These characteristics of the time-oscillation of 
the auto-correlation function suggest that the time-oscillating 
behavior of the auto-correlation function reflects 
a collective movement of the system.

\section{Damping Behavior of Momentum Auto-Correlation Function in a Long Time Interval} 
\label{EnvelopOscillation}

   In this appendix we discuss two points about 
the momentum auto-correlation function in the long time interval: 
(i) The shape of the envelop of the time-oscillation in 
the auto-correlation function 
$C_{x}$ for the $x$-component of the momentum, and 
(ii) The behavior of the auto-correlation function 
$C_{y}$ of the $y$-component of the momentum on a much longer 
time scale than that shown in the text of this paper. 

\begin{figure}[!htb]
\vspfigA
\caption{
      (a) Absolute values $|C_{x}|$ and $|C_{y}|$ of the 
   auto-correlation functions for the $x$-component 
   and the $y$-component of the momentum, respectively, 
   as functions of the collision number $n_{t}$ presented as 
   a log-log plot. 
      The solid line is a fit of the envelope of the 
   time-oscillating part of the auto-correlation 
   function $C_{x}$ to an exponential function.
      The dotted line is a fit of the auto-correlation 
   function $C_{y}$ to a $\kappa$-exponential function 
   (Eq. (\ref{kexponential})) 
   and the dotted broken line is a fit to 
   an exponential function in the region where 
   there is a deviation from the $\kappa$-exponential function. 
      (b) The time-oscillating part of the auto-correlation part 
   as a graph of the absolute value $|C_{x}|$ as a function 
   of the collision number $n_{t}$ presented as a linear-log plot. 
      The broken line is a fit to a sinusoidal function 
   multiplied by an exponential decay function 
   (Eq. (\ref{SinusoidalPower})), and the solid line is 
   its envelope, which is the same as the solid line in Fig. (a). 
      The system is a quasi-one-dimensional 
   system consisting of $50$ hard-disks with (H,P) boundary condition. 
   }
\vspfigB
\label{figG1tcfLongScaleCxCy}\end{figure}  

   Figure \ref{figG1tcfLongScaleCxCy} shows the absolute values 
of $x$ and $y$ auto-correlation functions $|C_{x}|$ ($|C_{y}|$) 
as functions of the collision number $n_{t}$ in a 
quasi-one-dimensional system of $50$ hard-disks 
with (H,P) boundary condition. 
   In Fig. \ref{figG1tcfLongScaleCxCy}(a) these graphs are plotted 
as log-log plots, while in Fig. \ref{figG1tcfLongScaleCxCy}(b) 
the graph for $|C_{x}|$ is plotted as a linear-log plot. 
   The collision number interval in this figure is about 
ten times as long as the previous ones in this paper, 
and we took a much longer time-average (eg. over $10^{9}$ collisions) 
to get this data. 

   As shown in Sec. \ref{MomentumAutoCorrelationFunctions}, 
the auto-correlation function $C_{x}$ for the $x$-component 
of the momentum decays exponentially initially.
   After the initial decay, the time-oscillating 
region of $C_{x}$ starts. 
   We fitted this region of $C_{x}$ to a sinusoidal function 
multiplied by an exponential function, namely Eq. 
(\ref{SinusoidalPower}), with the fitting parameters 
$\mathcal{A} \approx  0.0402$, $\beta' \approx 0.000368$,  
$T_{acf} \approx 2.03\times 10^{3}$ and $\xi \approx 1.53$ 
as the broken lines in 
Fig. \ref{figG1tcfLongScaleCxCy}(b). 
   The solid lines in Fig. \ref{figG1tcfLongScaleCxCy}(a) and (b) 
are the envelope $y = \mathcal{A} \exp\{-\beta' x\}$ of this function. 
   In order to see its exponential behavior 
we show, in Fig. \ref{figG1tcfLongScaleCxCy}(b), 
the linear-log plot of $|C_{x}|$ for the time-oscillating region 
of $C_{x}$, in which the exponential decay is represented as 
a straight line. 
   In this linear-log plot  
the local maximum points of $|C_{x}|$ are clearly 
on a straight line.    

   In Sec. \ref{MomentumAutoCorrelationFunctions} we 
also showed that 
the auto-correlation function $C_{y}$ for the 
$y$-component of the momentum is nicely fitted to 
a $\kappa$-exponential function (\ref{kexponential}). 
   This is also shown in Fig. \ref{figG1tcfLongScaleCxCy}(a) 
as the fit line to the $\kappa$-exponential function 
with fitting parameter values 
$\alpha'' \approx 0.00746$ and $\kappa \approx 1.48$ 
(dotted line).  
   However, Fig. \ref{figG1tcfLongScaleCxCy}(a) shows that 
there is a deviation from this functional form
on a long time scale. 
   Such a deviation is significant in the collision number region 
$n_{t} >  10000$ in this graph. 
   We fitted the auto-correlation function $C_{y}$ to 
an exponential function $y = \mathcal{A}'\exp\{-\alpha''' x\}$ 
with fitting parameter values $\mathcal{A}' \approx 0.0369$ 
and $\alpha''' \approx 6.25\times 10^{-5}$
(the dotted-broken  
in Fig. \ref{figG1tcfLongScaleCxCy}(a)) 
in the region where $C_{y}$ deviates from the $\kappa$-exponential.

%

%
%

\begin{thebibliography}{99}
\bibitem{noteC1}
      It should be noted that the auto-correlation function 
   $C_{lya,x}^{(2N-3)}$ includes an effect due to the longitudinal 
   Lyapunov mode without momentum (the second term 
   on the right-hand side of Eq. (\ref{LyapuOscilWPx1})) as 
   well as the momentum proportional term  
   (the first term on the right-hand side of 
   Eq. (\ref{LyapuOscilWPx1})). 
      On the other hand, the assumption (\ref{acflyavecexpl}) 
   is for the term for the momentum proportional Lyapunov mode 
   only. 
      In this sense, Fig.  \ref{figC6TcfLyaVec} does not represent
   a proof of the assumption (\ref{acflyavecexpl}).
\end{thebibliography}
\end{document}